%% file: main.tex
\documentclass[compsoc,conference,a4paper,10pt,times]{IEEEtran}
\IEEEoverridecommandlockouts

\usepackage{graphicx}
\usepackage{amsmath}

\usepackage{caption}
\usepackage{subcaption}
\usepackage{float}
\usepackage{soul}
\usepackage{xcolor}

\usepackage{array} 
\newcolumntype{R}[1]{>{\raggedleft\arraybackslash}p{#1}}

\usepackage[absolute,showboxes]{textpos}
\usepackage{hyperref}
\usepackage{cleveref}

\usepackage{cleveref}

\usepackage{seqsplit}
\newcommand{\hash}[1]{{\ttfamily\seqsplit{#1}}}

\usepackage{enumitem}

\usepackage{pifont}
\newcommand{\yes}{\ding{51}}%
\newcommand{\no}{\ding{55}}%

\usepackage{xurl}

\begin{document}

\setlength{\TPHorizModule}{\paperwidth}
\setlength{\TPVertModule}{\paperheight}
\TPMargin{5pt}
\begin{textblock}{0.8}(0.1,0.02)
     \noindent
     \footnotesize
     If you cite this paper, please use the EuroS\&P reference:
     Zeya Umayya, Manan Aggarwal, Manan Chugh, Mann Nariya, Yogesh Kaushik, and Sambuddho Chakravarty. The Invisible Ink of the Android Malware World: A Longitudinal Study on the Usage of Covert Communication Channels. 2026 IEEE 11th European Symposium on Security and Privacy (EuroS\&P). IEEE, 2026.
\end{textblock}

\title{The Invisible Ink of the Android Malware World: A Longitudinal Study on the Usage of Covert Communication Channels}


\author{
\IEEEauthorblockN{
Zeya Umayya,
Manan Aggarwal,
Manan Chugh,
Mann Nariya,
Yogesh Kaushik,
Sambuddho Chakravarty
}
\IEEEauthorblockA{
\textit{IIIT Delhi, India} \\
Email: \{zeyau, manan22273, manan21335, mann22278, yogesh20163, sambuddho\}@iiitd.ac.in
}
}
\maketitle
\begin{abstract}

Proxies, VPNs and Tor have long helped the privacy community and users in censored regions to fight censorship. However, the same tools can be maliciously exploited by malware and botnets to conceal their communication to external command and control servers. Despite being a critical concern fueled by the proliferation of malware based attacks, no longitudinal studies have analyzed how malware applications use covert channels (CC) to evade detection.

We fill this gap by performing the first study of the usage of covert channels in the Android malware ecosystem. To that end, we develop a complex multistage pipeline that combines static and dynamic analysis to investigate both system and 
network-level features. We applied this pipeline on a corpus of 3.5M Android malware spanning 2009 to July 2025 (one of the largest sample for such a study). Our carefully crafted static validation rules uncovered 288K APKs that used CCs spanning 511 malware families and CC usage growing exponentially from 0.30\% (2012) to 50\% (2025). The dynamic execution of a subset of these applications on physical mobile phones not only confirmed the results of static analysis based results but also enabled a deeper analysis. Overall, we identified 19,308 unique IP addresses being contacted in 85 countries, out of which we were able to explicitly validate the presence of CCs for 59 IP addresses across 17 countries. Further, we performed a longitudinal dataset study spanning over 16 years for CC based malware and found that CC usage has evolved, \textit{e.g.,} some malware adopted by using more than one CCs (one family used more than 5 CCs); others switched between them periodically (one family switched CC usage 40 times from 2019 to 2025).

Lastly, we tested two state-of-the-art malware detection solutions on CC-based malware identified in our work. We found that they fail miserably, yielding high false rates ($>$ 50\%), underscoring the need to consider the effect of CCs on malware detectors.
Overall, we believe that our work advances the community's understanding of CC based malware and lays the groundwork for future studies to consider CC usage when developing malware detectors.

\end{abstract}


\input{sections/intro}
\input{sections/background}

\input{sections/pipeline}
\input{sections/measurement_results}

\input{sections/discussion}

\input{sections/related_work}

\input{sections/conclusion}
\input{sections/acknowledgment}

\bibliographystyle{IEEEtran}
\bibliography{ref}

\appendices
\input{sections/appendix}

\end{document}

%% file: sections/intro.tex
\section{Introduction}
\label{sec:intro}

Often malware use tools like \textit{Tor} \cite{dingledine2004tor}, its \textit{bridges} \cite{umayya2023ptperf}, \textit{I2P} \cite{i2p} \textit{VPNs} \cite{vpn} and \textit{proxies} to hide communication with Command \& Control (C2) servers or peer bots. Multiple articles indicate that alongside desktop malware \cite{domain_fronting_usage,tor_usage,covert_usage,dark_utilities_usage1,dark_utilities_usage2},
Android malware increasingly employ such mechanisms ~\cite{blog1_tor,blog4_proxy,blog5_tor,blog2_proxy,blog5_proxy}. We collectively refer to such tools as \textit{covert channels (CC)}. 

While there exists work on studying CCs (VPNs/Tor/PTs/Proxies/I2P), there is a clear research gap on investigating and characterizing Android malware that uses CCs. 
Further, it is unclear if current state-of-the-art malicious traffic detection methods\footnote{These solutions are positioned mostly at edge router observing organizational traffic at a large scale.} effectively identify such APK traffic. Our preliminary evaluation in \Cref{sec:mal_sota_detection} suggests these methods often fail to detect malware APKs that employ CC tools/packages.
Thus, there is an urgent need for a dedicated study to find and investigate such malware and their evolution.



\noindent \textbf{Challenges and missing ground truth}:
Identifying APKs that use CCs across millions of malware samples is challenging on multiple fronts, as evidence of their usages are currently fragmented across disparate blogs, reports by security researchers and industry analysts.
Firstly, the hashes cited in existing blog posts seldom point to the actual APKs
employing CCs \cite{blog1_tor,blog2_tor}. Instead, they usually correspond to payload or ELF executables embedded
within malware. 
Thus to the best of our knowledge, no curated datasets are publicly available to academics, or prominent security symposia like Defcon/Blackhat \cite{blackhat1,defcon1}.
Secondly, while VirusTotal (VT) \cite{vt_api} has long served as a reliable and de-facto source for obtaining generic malware related information and heavily cited in prior works \textit{e.g.,} \cite{arp2014drebin,xu2018deeprefiner,wu2019malscan,dodia2022exposing,sebastian2020avclass2,alecci2024androzoo}, it does not explicitly identify the use of such communication tools.
Existing work over third party libraries \cite{wu2023libscan,zhan2021research,zhang2020empirical}, mostly look for a collection of general libraries found in malware APKs and does not pin-point towards CC usage.
Thus we do not have ground-truth of such malware APKs to begin with. 
Further, dynamically executing every malware APK available today to observe CC creation is infeasible at scale. 
Moreover, there is a lack of methods to check CC creation at runtime. Thus such complexities are an imperative to explore novel methods involving static analysis, which can scale 
rapidly and cover millions of malware.



Thus to address such issues, we develop a framework that first identifies Android malware that use CCs, and then 
analyzes them using static code inspection and selective dynamic execution. 
Our contributions are presented under three components: static analysis, dynamic analysis, and the characterization of the collected data.

\noindent \textbf{Static analysis:}
The purpose of performing static analysis is to obtain indicators of CC usage in the form of specific rules, which we term \textit{static validation rules (SVR)}. 
Due to the absence of ground truth, we adopted an exploratory approach to identify and characterize such malware.
To collect a seed list of APKs, we look for potential keywords related to each CC in malware VT reports. 
Since it is not known how malware creators may have incorporated those CCs, we 
search for CC usage indicators in seed malware APKs, following how they are usually integrated in their benign counterparts. 
Since each CC demands a very different type of requirement and setup to work, their indicators of CC usage require dedicated efforts respectively.
After each CC has been studied, collected information is converted into SVRs and then it is applied on millions of malware in an automated manner.


As a \textbf{first contribution}, we explain how we design SVRs for each CC. To refine our approach, we analyzed how CC tools are integrated into third-party APKs by manually reviewing relevant documentation on the former's websites.
For Tor/TorPT/I2P, we identified Android packages (libraries) used for integration, making their presence a strong indicator of Tor/I2P usage. 
Other relevant rules include the presence of Tor/I2P binary identified upon static inspection of malware APKs. 
For VPNs, we identified Google Playstore's VPN APKs
and decompiled them to determine package used. 
Unlike most VPNs, proxy APKs in app stores usually have other functionalities as well (\textit{e.g.} doubling up as VPNs). 
Thus, we first aggregated all possible provider names from prior research works \cite{bian2022shining,mehanna2024free,mi2019resident,mi2021your}, and inspected their APKs to identify packages\footnote{Packages have downloads in millions , see \cref{sec:vpn_pkg,sec:proxy_pkg}.} specifically used for proxy functionality (\Cref{sec:pa_static_validation}). Additionally, we verified runtime usage of proxy/VPN package methods by instrumenting and dynamically executing 50 APKs (see \Cref{sec:apk_instrumentation} for details). 


Since APKs are often \textit{obfuscated}~\cite{duan2018things,siddiqui2024overview}, SVRs may not always apply. Additionally, if malware creators have modified the original CC packages, then our SVRs may fail.
To ensure we didn't ignore such cases, we try to find similarity of already identified CC APKs at two levels---firstly matching CC library using \verb|LibScan|~\cite{wu2023libscan} to determine if a given APK has a particular library packaged in it (even for obfuscated APK). 
And secondly, matching two APKs entire code using \verb|MatchScope| \cite{feng2024accurate} where the first is a validated malware APK and the second is a non-validated one (for results see \Cref{sec:matchscope,sec:obfuscation_handling}). 
Although \texttt{LibScan} and \texttt{MatchScope} are state-of-the-art tools, they are still far from handling all types of obfuscations (as described ahead in \Cref{sec:limitations}).



\noindent \textbf{Dynamic analysis:}
Thus, at the end of the static validation process, we have a set of potentially covert malware that use Tor/PTs/I2P/VPNs/proxies as CC.
As a \textbf{second contribution}, we develop dynamic validation rules to validate potentially covert malware. Dynamic analysis involves executing malware APKs either on an emulator or on a real device. 
It is widely known that the dynamic analysis of malware often falls short of full success~\cite {ruggia2024unmasking,mbc}.
For example, malware attempts to bypass the underlying test environment, either slowing down its functionality or completely masking its true nature due to the absence of a live C2 or its peers.
To overcome these challenges, we set up physical devices to run malware and provide a 5-minute window for malware to create CC and exhibit its true behavior, while also not extending the period beyond this, in accordance with ethical practices (see \Cref{sec:ethics}).



To develop dynamic validation rules, we examined evidence at both network and system levels. Network traces helped identify Tor relays and detect VPN traffic using regex-based \cite{maghsoudlou2023characterizing} and nDPI \cite{deri2014ndpi} analyses across multiple protocols (\textit{e.g.,} OpenVPN, WireGuard). System-level inspection of logs and process traces revealed that each CC generated unique, identifiable patterns, forming the basis of our dynamic validation approach.

\noindent \textbf{Discovery and characterization of covert malware:}
After both static and dynamic validation processes, we perform PCM characterization based on their presence across years, their evolution, and their static and dynamic behavior, as our \textbf{third contribution}. 
While considering 3.6M malware APKs (with VT-score $>$ 0) from AndroZoo \cite{androzoo}, between 2009 and 2025 (16 years data), we applied both SVRs on APKs and keyword searches on VT reports. 
Overall, for 3.6M malware, keyword searches over their VT reports flagged 102k APKs, where 81k were false positives, while our static validation rules over these APKs alone discovered 288k APKs.
Thus, our static validation pipeline flagged approximately \texttt{288k} APKs using some CC:
150 (Tor), 3.5k (TorPT), 2.4k (I2P), 17k (VPN), and 271k (proxies).
It contains 511 distinct families overall, with 12, 39, 23, 127, and 477 families that use Tor, TorPT, I2P, VPN, and Proxy, respectively.
Additionally, 215k singleton samples were also identified. It has 124k unique APKs based on their package names. Proxy has more unique APKs (119k) than Tor, which has the least number of unique APKs (80). 
To further trace back the developers of these malware and find other such malicious APKs from the same developers, we statically inspected their certificates.  Interestingly, it revealed widespread use of common certificate keys and fake organization names across thousands of APKs (see \Cref{sec:apk_certs}).

We also analyzed the evolution of malware APKs and their families to understand how they transitioned between different covert channels over time. Our dataset, spanning 2009–2025, offers a broad temporal scope to capture and study these shifts in communication strategies.
We found numerous instances where the same malware APKs 
transitioned from using one CC to multiple. 
There were such malware that continuously changed their CCs. It may be to hide and change the signature of their APK to avoid anti-virus engines that detect APKs using their hash values (\Cref{sec:evolution}). 

\textit{Dynamic Results:}
For dynamic level validation and characterization, we executed randomly sampled 13.7k malware (out of the 288k) APKs across all categories on Google Pixel-5 phones\footnote{Refer to \Cref{sec:ethics} for ethics consideration.}.
We didn't provide any user input apart from allowing permission to create VPN tunnels (since the purpose is to observe covert behavior). 
Highly sophisticated malware often bypasses dynamic execution to conceal its true behavior \cite{ruggia2024unmasking,mbc}. Despite the inherent limitations of dynamic analysis, we conducted this experiment as a pragmatic necessity to accurately study malware behavior.
Overall, it resulted in 6.4k successful runs, out of the 13.7k APKs. 
\Cref{fig:succes_rate_dynamic} shows that APKs from 2019 onwards had more than 40\% of execution success rate.
Though 13.7k APKs appear less, they span various periods and are executed on real devices, unlike studies using emulators (Mercaldo \textit{et al.} \cite{mercaldo2023explainable}). 
Upon analyzing the dynamic execution data, we observed that malware try connecting to Tor, VPN and proxy servers spread across the globe in countries like China, USA, Germany, France \textit{etc.} (\Cref{sec:dyn_analisis}).
We also observed that some malware use the same IP endpoints for connecting to their CCs. Such trends may shift with longer analysis.

Overall, we make the following contributions:
\begin{itemize}[leftmargin=*]
    \item  We conduct the first systematic investigation of Android malware leveraging covert channels (CCs) such as Tor, Tor pluggable transports, I2P, VPNs, and proxies, highlighting gaps in existing detection approaches.
    \item We design and implement a novel static analysis framework based on carefully curated SVRs to identify indicators of CC usage in APKs, enabling scalable analysis across millions of samples despite the absence of ground truth. 
    \item We develop dynamic validation rules based on network- and system-level artifacts, and validate malware behavior through large-scale execution on physical devices, capturing realistic CC usage patterns.
    \item \textit{Large-scale dataset and characterization:} Applying our pipeline to 3.6M APKs, we identify $\approx$288K malware samples across 511 families using covert channels, and perform in-depth characterization of their distribution, evolution, and behavior over 14 years.
\end{itemize}

We have released the resulting dataset, comprising malware hashes and accompanying analysis code, through a publicly available GitHub repository~\cite{ink_github} to facilitate further dynamic and static analyses.
Overall, we believe that our work advances the community's understanding of sophisticated malware enabling studies over Android malware evolution, shedding light on how adversaries continuously adapt and refine their malicious designs.


%% file: sections/background.tex
\section{Background}
\label{sec:background}

\subsection{Covert Channels (CC)}
\label{sec:covert_channel}
There are many ways of communication over the internet without revealing one's digital identity (particularly IP address) 
such as Tor, VPNs, proxies, pluggable transports and I2P.
We provide a brief of each CC.

\noindent \textbf{Tor}: It was proposed in 2004 as an anonymous routing tool \cite{dingledine2004tor}. 
Its network comprises relays, running as proxies, usually known as \textit{onion routers}. 
A Tor client communicates with a server (\textit{i.e.}, destination) by relaying traffic via a cascade of relays, called \textit{guard}, \textit{middle} and \textit{exit}, collectively referred to as a \textit{circuit}. 
Tor is widely used as a censorship circumvention tool. The client to relay traffic reveals no information about the actual destination of the messages.


\noindent \textbf{Pluggable Transports (TorPT)}: Tor relay IPs are publicly advertised through \textit{directory services}~\cite{dir}. Often relay IPs are blocked by censors. To bypass the censor, TorPT relay Tor traffic through alternate methods such as tunneling (Dnstt, WebTunnel \textit{etc.}), proxying (Snowflake, Meek \textit{etc.}), mimicking other protocols (Cloak, Marionette \textit{etc.}) and wrapping data in new encrypted protocols (Obfs4 \textit{etc.}) \cite{umayya2023ptperf,khattak2016sok}. This way a censor cannot spot Tor relay IPs in flows and thus may not attempt to block them. 
These are sometimes known as bridges. 
We will refer to these as TorPT throughout the paper. 

\noindent \textbf{VPN}: Virtual Private Networks (VPN) \cite{scott1999virtual} were created to virtually connect to the private address spaces. VPNs tunnel the traffic of such address in a manner that an observer only sees the public IPs of the tunnel end-points, and not those of the private networks. A VPN client usually establishes such tunnels with VPN servers/gateways, using popular protocols like \textit{OpenVPN}~\cite{openvpn}, \textit{Wireguard}~\cite{donenfeld2017wireguard} \textit{etc.}
Due to their similarity with proxies, they are also used to evade censorship by hiding the actual destination IP from an eavesdropping adversary.

\noindent \textbf{Free Proxies}: Free proxies have always attracted more users, because they are free and often faster than VPNs \cite{proxy_vs_vpn}. 
These are most commonly used for bypassing censorship and geo-blocking.
They commonly support HTTP and SOCKS protocol. HTTP proxy is used for accessing web content and SOCKS by default supports multiple applications over it \cite{mehanna2024free}.  

\noindent \textbf{I2P}: I2P~\cite{i2p},
akin to Tor hidden service, ensures two-way anonymity \textit{i.e.,} \textit{initiator} and \textit{responder}, using \textit{garlic routing} \cite{garlic_routing} to exchange messages between nodes. It establishes inbound and outbound tunnels between I2P nodes, re-established every 10 minutes. Its first mobile application was released for Android in 2011~ \cite{i2p_tags}.


\subsubsection{Exploitation by malware}
\label{sec:back_covert}
All the above CCs are also used by malware to hide the network communication to end-points such as C2 servers or peer bots. As per the reports by multiple security researchers \cite{domain_fronting_usage,tor_usage,covert_usage,dark_utilities_usage1,dark_utilities_usage2} malware have started using Tor, domain fronting (also used by meek and snowflake \cite{snowflake,meek}) for stealthy C2 communications.\\


\noindent \textbf{Specific to Android:}
As per multiple reports \cite{blog1_tor,blog2_tor,blog4_proxy,blog1_proxy,blog3_proxy,blog5_proxy}, Android malware also use such CCs for hiding their traffic. Some famous malware families such as \verb|Hydra|, \verb|Gafgtyt| \textit{etc.} have been found using Tor to hide their C2 communications. It is a complex process to integrate covert communication tools into an APK, other than the original one (\textit{i.e.}, the covert application's) its developers envisaged. There are many ways to integrate CC functionalities
into an APK. These include using libraries (or packages), or by directly loading compatible binaries of available covert communication applications like I2P, Tor and TorPTs.
The complications arise due to how malware creators implement the functionality of covert communication into APKs. Intuitively simpler integration options may lead to more malware APKs using them. 
Generally, malware creators leverage the development efforts by the providers of such CCs, which were originally meant for benign APKs.

\subsection{APK Analysis Challenges}
\label{sec:mal_analysis}
No standard methods exist to integrate covert communication functionality into APKs, making it hard to detect such channels in their source code. One approach is to run APKs on phones and monitor for CC creation, but dynamically analyzing millions of malware is infeasible and time-consuming with execution challenges. Instead, we used faster static analysis to reduce the search space of approximately 4M APKs.

\begin{figure*}[t]
	\centering
	\includegraphics[width=0.94\textwidth]{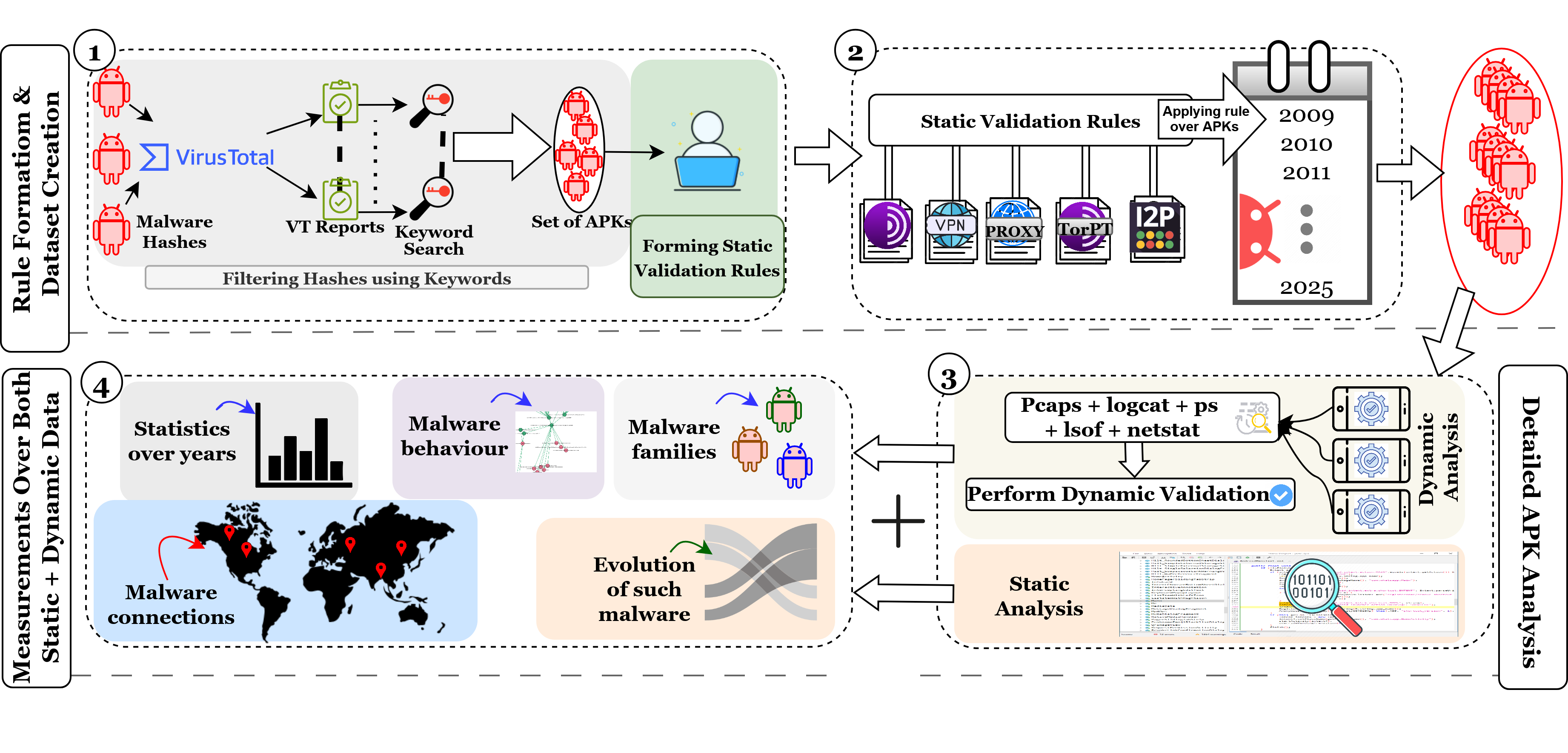}
    \caption{Complete Work Pipeline. \textcircled{1}: Formation of static validation rules that can be applied on APKs using static analysis over them. \textcircled{2}: Application of channel's (Tor/VPN/Proxy/TorPT/I2P) rules over APKs from 2009-2025. \textcircled{3}: Static and dynamic analysis on the dataset. \textcircled{4}: Results analysis and presentation of the measurement results.}
	\label{fig:pipeline}
\end{figure*}
\subsubsection{Static analysis challenges}
\label{sec:static_ch}
Although static analysis is more efficient compared to dynamic, but it has its own challenges. \textit{E.g.}, while searching for the presence of CC code, it is not known prior whether the APK is obfuscated or not. Apart from \texttt{APKiD} tool \cite{apkid_blackhat}, there is none other that can detect APK obfuscation. But this also has its limitations, making it unsuitable for majority of the APKs. 
Thus, for multiple APKs, static analysis does not result in successful identification of the CC, and such APKs are skipped from further analysis even though they might be using such mechanisms. 
As a solution to detect obfuscated libraries or code, we use state-of-the-art tools such as LibScan \cite{wu2023libscan} (to match library) or MatchScope \cite{feng2024accurate} (to match entire code). However, these tools can handle only a few types of obfuscation. Thus, our final dataset will represent a lower bound of the number of CC APKs that might be present in the wild.



\subsubsection{Dynamic analysis challenges}
\label{sec:dynamic_ch}
Our work also requires dynamic analysis of APKs for validating if they are indeed using CCs, and also to observe their runtime behavior. There are different challenges with each type of CC. \textit{E.g.}, some of the VPN-based APKs require user input before establishment of the VPN tunnels. Thus, we need to automatically detect such pop-ups and grant the necessary permissions. Other high profile malware corrupt the operating system itself, and thus leave the device stuck in an unknown state that prevents dynamic analysis. In such extreme cases, manual intervention is required to reset the device by debugging the issues at hand. 
Although we run malware APKs on mobile phones to observe CC creation, the former may not always create them immediately upon execution. As a result, detecting CC connection attempts within the limited duration of dynamic analysis might not always be possible.


%% file: sections/pipeline.tex
\section{Proposed Approach}
\label{sec:proposed_approach}


\textbf{Overview:}
We provide an overview of the complete pipeline in \Cref{fig:pipeline} to discover malware that use CCs for communication. 
Step \textcircled{1} involves inspection of APKs filtered using specific channels' keywords on VT reports, so as to determine what heuristics/rules that are required to identify such channels.
Step \textcircled{2} applies these rules on all APKs from 2009 to 2025 (\Cref{sec:pa_hashes}). It involves statically analyzing APKs and then applying the rules to them, to spot the use of CCs. These APKs are then collected into a new dataset. 
Step \textcircled{3} takes a set of the APKs from this dataset and performs dynamic analysis by running them on real phones. The purpose is to first dynamically validate if CCs are actually established, and secondly to study the runtime behavior of such malware APKs. 
Additionally, it also involves static analysis of the entire dataset for determining the behavior of these malware especially with respect to how and when they use the CC. 
The outcomes of third step are combined and fed to step-\textcircled{4}. This final step performs different types of measurement analysis (detailed in later sections) and presents insightful statistics and behavioral results. 
We first provide details about the malware we considered for our study and then move through each step in detail.

\subsection{Android Malware Collection}
\label{sec:pa_hashes}
\Cref{table:initial_dataset} presents the statistics of our dataset, which includes malware APKs from the years 2009 to 2025\footnote{Data collected till July month for 2025.} (Android OS was released in 2008).
We consider the maximum possible time duration to study the evolution of such malware over the years.
For APK downloads, we used \verb|AndroZoo| repository \cite{androzoo} containing millions of APKs. 
For each APK hash having VirusTotal (VT) score of atleast one (as provided by the meta-data file in AndroZoo), we also download its VT report using VirusTotal API. 
For this we subscribed to VT Academic API \cite{vt_api} which provides more API requests per day than in general. VT reports, updated dynamically \cite{vt_howto}, may differ from AndroZoo’s snapshot-based reports. Thus, we used the latest VT reports as ground truth.

\begin{table}[t] 
\centering
\footnotesize{
     \caption{Yearwise count of APKs. Total available APKs: 24.5M ; Total number of APKs with VT-Score $>$ 0: 3.6M }
      \label{table:initial_dataset}
    \begin{tabular}{R{0.5cm}| R{1.15cm}| R{1.15cm}| R{0.5cm}| R{1.15cm}| R{1.15cm}}
    
        \hline
        \textbf{Year} & \centering \textbf{All Available} \arraybackslash  & \centering \textbf{VT-Score $>$ 0} \arraybackslash &
        \centering \textbf{Year} \arraybackslash & 
        \centering \textbf{All Available} \arraybackslash & 
        \centering \textbf{VT-Score $>$ 0} \arraybackslash \\ \hline
        2009 & 1 & 0 & 2010 & 15 & 0 \\ \hline
        2011 & 4169 & 32 & 2012 & 60125 & 9970 \\ \hline 
        2013 & 618951 & 219123 & 2014 & 1614685 & 541250 \\ \hline
        2015 & 508419 & 131981 & 2016 & 2308774 & 606736 \\ \hline
        2017 & 561345 & 123704 & 2018 & 2146286 & 350343  \\ \hline
        2019 & 2034351 & 432544 & 2020 & 3014674 & 269584 \\ \hline
        2021 & 3538073 & 342327 & 2022 & 4862188 & 386492  \\ \hline
        2023 & 2603080 & 88738 & 2024 & 636821 & 25399 \\ \hline
        2025 & 132274 & 2454 & \textbf{Total} & \textbf{24511957} & \textbf{3652832} 
        \\ \hline

    \end{tabular}        
    }

\end{table}

\subsection{Formation of Static Validation Rules}
\label{sec:pa_static_validation}

For this work, we considered popular CCs such as Tor, VPNs, proxies, I2P, and TorPT being used for hiding the communication with command and control servers mostly. 
Next, we next will explain step \textcircled{1} in detail.

\subsubsection{Filtering hashes using keywords}
\label{sec:step_1}

The purpose of forming keywords is to search them in VT reports. All VT reports with such keywords may or may not be actually related to the respective CC. Formation of relevant keywords for each CC requires a thorough understanding of each channel and its usage styles in multiple platforms such as Linux or Android. For each CC, we formed some keywords generally referring these. We provide those keywords in \Cref{table:keyword_table} (in \Cref{sec:keywords}).

As per \textcircled{1}, we initially generated VT reports for malware hashes and then applied keyword filtering over it. 
Further, we manually analyzed each report for false positives while searching for valid reports. After extensive efforts, we came across some APKs actually using Tor, VPN, proxy, TorPT or I2P. Upon statically analyzing them, we found some examples of such CCs being used in APKs. 
Overall, VT keyword filtering mostly generated false positives (details in \Cref{sec:keyword_filtering}). Thus, we conducted deeper
static analyses of the APKs to find indicator of usages related to CCs.


\subsubsection{APK inspection to create rules}
\label{sec:step_1}

While searching for keywords in VT reports, we observed that most of the matches led to false positive samples. Thus we concluded that searching keywords in VT reports is not effective in discovering such APKs. 
Thus, we looked for additional ways by which APKs could be using CC through the latter's official webpages. 

\textbf{For Tor and TorPT}, we found multiple possible ways to integrate them to an APK. These are listed as follows:
\begin{enumerate}[leftmargin=*, itemsep=0pt, topsep=0pt]
    \item Presence of Tor binary at \textit{res/raw/tor} or \textit{assets/tor}
    \item Presence of libTor.so, tor.so or torrc configuration files
    \item Presence of packages such as 
        \begin{enumerate}
            \item \textit{com.netzarchitekta.IPtProxy},
            \item \textit{info.guardianproject.onionkit},
            \item \textit{info.guardianproject.netcipher},
            \item \textit{org.torproject.jni.TorService}, and
            \item \textit{org.torproject.iptproxy.IPtProxy}
        \end{enumerate}
\end{enumerate}

\noindent These packages should never be the same as the main package of the APK, because that only happens for APKs of these CCs themselves. Such condition makes sure that we are not flagging the original CC channel APKs themselves.

\textbf{For VPN}, while analyzing filtered APKs, we found different packages in each APK, and a specific Android permission (\texttt{BIND\_VPN\_SERVICE}\footnote{Introduced in 2011 for Android v4 and above.}) required to establish tunnels.
Unlike Tor, there are many VPN service providers. Thus, it is not trivial to fetch the list of all packages providing the functionalities of a VPN service. 
Since we were unsure about whether the packages that we found in VT reports were related specifically to VPNs, we took a different path and hypothesized that malware creators maybe re-using existing VPN APKs' packages available through official stores. 
For this, we scrapped through Playstore, F-droid, and Github, first to fetch all VPN APKs through keyword search and discovered that their packages carry out the tasks of VPN management. We found 49 such APKs and thus 49 unique packages. We provide their details in \Cref{sec:vpn_pkg}, showing millions of downloads per APK. We used these as a seed list. 
For a final confirmation regarding permission, we also thoroughly inspected a benign APK about how a VPN connection is made and what permissions are required for it to function properly. We finalized that along with a VPN package the APK under test must also mention \texttt{BIND\_VPN\_SERVICE} in its manifest file. The permission or the package alone are not sufficient since many malware creators add unnecessary permissions to confuse malware detectors trained on such features \cite{sun2023mate}.

Similar to VPNs, \textbf{for proxy}, VT reports were not very helpful either. Thus we took a different route. We looked for prior research on free proxies in Android OS \cite{mi2019resident,bian2022shining,mehanna2024free,mi2021your}. We observed that proxy APKs are also usually distributed through their providers. On Playstore, we did not find many APKs with only proxy functionality. For a clean SVR of proxy usage, we first required to have APKs providing only proxy as a service, and nothing else. As a result, we decided to find such APKs on proxy providers' websites. 

Thus we unified all 58 proxy providers names from multiple sources \cite{mi2019resident,choi2020understanding,mi2021your,mehanna2024free}. We collected 10 APKs that were mentioned on the websites of these developers. The purpose was to find out how the proxies are implemented 
in a benign APK. After decompilation and inspection of the APK code we found that these APKs themselves use third-party packages. For completeness, we also considered all previous versions of these APKs which amounted to be 165 in total. 
Specifically, each APK was decompiled and analyzed to extract its package structure, to identify newly introduced or third-party packages. We then examined the usage of these packages within the application codebase to determine whether they were actively invoked and whether their functionality aligned with proxy CC behavior.
Through this process, we identified a total of 15 distinct third-party packages that implement proxy or CC-related functionality (listed in \Cref{sec:proxy_pkg}). These packages were observed across an initial set of 10 APKs, with additional libraries emerging through version-wise analysis.

\textbf{For I2P}, we inspected benign and VT-filtered APKs to craft static validation rules, and 
four integration methods:
\begin{enumerate}[leftmargin=*, itemsep=0pt, topsep=0pt]
    \item Presence of I2P .so files such as \hash{libjbigi.so}, \hash{libi2p.so}, \hash{libi2pd.so}, and \hash{libinvizible.so} in \textit{lib} folder.
    \item Configuration/help files such as \hash{i2ptunnel\_config},  \hash{help\_i2ptunnel.html}, and \hash{i2pd.config} files in \textit{res/raw} folder, \hash{i2pcrt.crt} in assets folder.
    \item Presence of packages like: 
        \begin{enumerate}
            \item \textit{net.i2p.android.lib.helper},
            \item \textit{net.i2p.android.router.service},
            \item \textit{net.i2p.crypto.eddsa.math},
            \item \textit{net.i2p.data.i2cp}
        \end{enumerate}
    \item Presence of strings \textit{e.g.,} \hash{logger.record.net.i2p, i2np.ntcp.enable, i2np.udp.enable, router.inboundPool, router.outboundPool} in configuration files.
\end{enumerate}

\subsubsection{Special case: Obfuscated and modified library code}
\label{sec:libscan_related}

\noindent \textbf{Obfuscated libraries:} In Android, benign/malware creators obfuscate the APKs to generally hide their source code and impede correct decompilation \cite{ruggia2024unmasking}. Thus applying above static validation rules on such obfuscated APKs may not work. Thus for obfuscated APKs we devised another strategy. 
Similar to how we looked for the presence of specific libraries for each CC, we tried to find the libraries in obfuscated APKs. Instead of searching for the exact library name, we searched for the contents of the libraries present in a malware APK. For this we used Wu \textit{et al.}'s \texttt{LibScan} tool \cite{wu2023libscan}.

To get the source code of all possible versions of a covert library, we first applied the static validation rules on each APK and obtained the first set of CC APKs. Then, we extracted libraries used in them\footnote{We implemented the library extraction code ourselves as it was not present in LibScan's codebase.} and search them in other non-flagged APKs to find their presence in case the malware creators tried to hide them using obfuscation. LibScan works for six obfuscation types and thus we are also limited by this\footnote{Types: identifier renaming, code addition, dead code removal, package flattening/repackaging, string encryption, control-flow randomization, manifest transformation, and data alignment.}. 

\noindent \textbf{Modified code:} Malware creators modify the known libraries as per their needs, and also to bypass static analysis. Even after such modifications, the semantic meaning of the code should not change.
To find known modified libraries code, we employ \texttt{MatchScope} \cite{feng2024accurate} on non-validated APKs. This is done in two steps -- (1) applying it to known libraries' codes to group them into semantically similar codebases. This clusters the validated APKs into fewer groups, reducing the searches required to find the correct match between these groups and the non-validated APKs.  
Next, in (2) the non-validated APKs are matched against these groups, thereby helping spot the presence of CC usage in their modified codes.

Thus, in \textcircled{2}, we applied all SVRs over malware APKs from 2009 to 2025. At the end of static validation, we obtained the dataset of APKs having CC present in them. After this step, the dataset was passed to step \textcircled{3} for a detailed APK analysis along with dynamic validation.

\subsection{Formation of Dynamic Validation}
\label{sec:pa_dynamic_validation}

\noindent \textbf{Dynamic setup:} Static validation rules help discover APKs that use CCs for network communication. To find out if indeed these malware establish CCs, we first execute these on real Android devices for five minutes and observe their network behavior by capturing the traffic (\textcircled{3} in \Cref{fig:pipeline}). Standard practice for dynamic analysis is 1-2 min. (Kuchler \textit{et al.}  \cite{kuchler2021does} and Umayya \textit{et al.} \cite{umayya2024comex}). However, we extended it to 5 min. to provide enough time for the malware to create tor-circuits/vpn/proxy connections, which may take 2 min. 

We also store relevant system-level information about the events that occur while the malware runs (using \texttt{netstat}, \texttt{ps}, \texttt{lsof} and \texttt{logcat}).
The purpose of capturing dynamic data is first to verify the establishment of CC and then further study their runtime behavior. 

Unlike static, dynamic analysis may not always produce concrete results. 
Not all malware APKs will try to establish CC connection immediately after installing and executing. It maybe the case that they require CCs at particular events, and not always. Next, we explain how we form the dynamic validation steps for each CC.

\begin{figure}[t]
	\centering
	\includegraphics[width=0.48\textwidth]{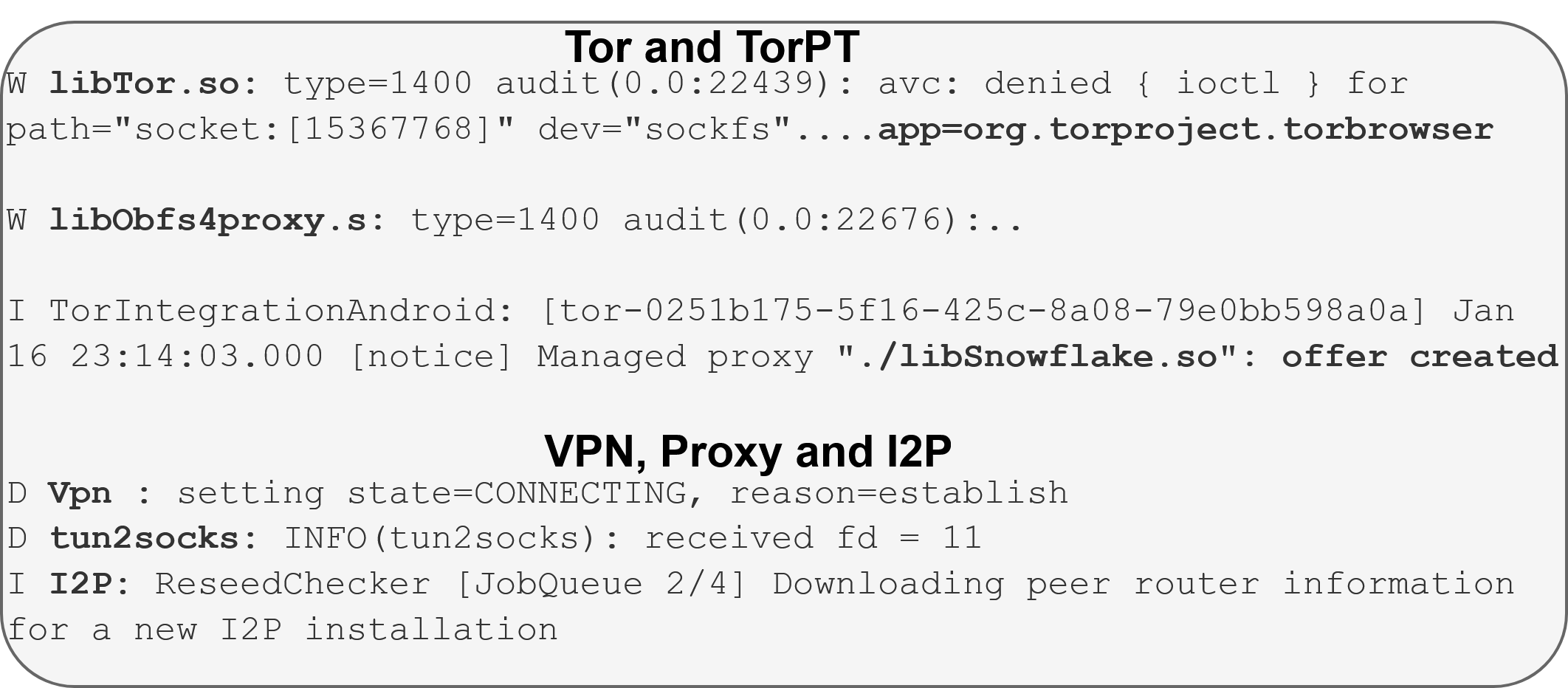}
	\caption{\texttt{Logcat} output in each covert APK.}
	\label{fig:logcat}
\end{figure}

\begin{figure}[t]
	\centering
	\includegraphics[width=0.49\textwidth]{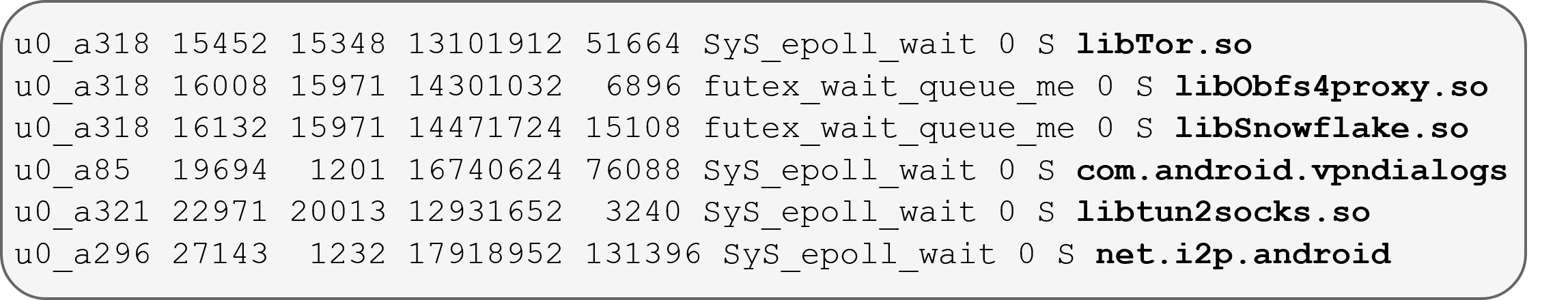}
	\caption{\texttt{Ps} output in each covert APK.}
	\label{fig:ps}
\end{figure}

\subsubsection{Tor validation}
\label{sec:tor_validation}
To connect to the Tor network, APK executes the Tor client program and downloads the necessary data by contacting the Tor network directory authorities. Once it bootstraps to 100\%, it creates circuits where the first communication end-point is a \textit{guard} node in the Tor network.

\textbf{Network-level validation:} Information about the running Tor relays is publicly announced by the Tor \textit{directory} servers \cite{tor_consensus}. 
These documents contain metadata about all currently reachable relays, including their IP addresses, ports, and roles (e.g., Guard, Exit). 
A client can fetch this information
from these servers using the \texttt{stem} library~\cite{stem}. Thus, every IP present in the pcap collected above can be checked against the list of guard nodes' IPs in Tor network's relays' information. 
Given the dynamic churn of Tor relays, we ensure that the consensus, data used to match relay IPs and ports, closely correlated to the time window during which the PCAP traces were collected. This reduces inaccuracies arising from such relay churn.


\textbf{System-level validation:} We can validate Tor execution through \texttt{logcat} logs as shown in \Cref{fig:logcat}.
This can also be found by running \verb|ps| command and looking for the \texttt{libTor.so} entry. Additionally, we fetch the data section of an APK under test, from the device. We check this
to see if there are files there that are populated/updated, and that may be necessary for Tor circuit creation. Once the APK is installed, it creates a directory named \texttt{app\_torfiles} with 7 files in it \textit{viz.} -- geoip, geoip6, lock, pid, state, tor, torrc. Their presence 
is a strong indicator that the installed APK is likely going to connect to Tor.  
We use this observation as a rule for dynamic validation.


\subsubsection{VPN validation}
To connect to a VPN server, an APK utilizes the VPN API provided by Android OS and requires \hash{BIND\_VPN\_SERVICE} permission. Other methods include adding native libraries and third-party SDKs.

\textbf{Network-level validation:} By capturing the network traffic, we can verify if an end-point IP:Port pair belongs to a VPN. To do this we relied on methods suggested
by Maghsoudlou \textit{et al.}  \cite{maghsoudlou2023characterizing}. In their approach, the authors send crafted probe packets, corresponding to different VPN services (OpenVPN, WireGuard, PPTP,
and SSTP), to IP:port pairs
they may wish to check. A legitimate response, determined by observing specific patterns in the probe responses, indicate that the end-point runs a VPN service. In our approach, we also
inspected the response packets, corresponding to an APK's communication, for similar patterns. We enlist these patterns in \Cref{sec:pam_paper}.

Apart from above method, we also used another tool \texttt{nDPI} (v4.12) \cite{deri2014ndpi} to find out the protocol used. It can detect 15 different types of VPN protocols (IPsec, GRE, PPTP, OpenVPN, CiscoVPN, WireGuard, FortiClient, Softether, CloudflareWarp, ProtonVPN, NordVPN, SurfShark, CactusVPN, TunnelBear, and OperaVPN). 

\textbf{System-level validation:} The \verb|logcat| and \verb|ps| commands log the creation of VPN connection establishment and trials. \Cref{fig:logcat,fig:ps} shows the statements that usually appear in logs.
While looking for new files created in the data folder, we observed one JSON file with a random string in its name\footnote{We found it in many APKs with fields similar to a VPN auth-token.}. 

\subsubsection{Proxy validation}
There are many ways in which an APK can connect to either a SOCKS/HTTP proxy server. These include OkHttp Android client utility, creating sockets using the Android class, \verb|proxy|, using VPN API of Android, or using third-party proxy libraries. Thus, for each proxy connection method, there would be a ``footprint'' in dynamic analysis data.

\textbf{Network-level validation:} We again used passive method and searched for the type of protocol of the APKs communication, using \texttt{nDPI} tool. It supports four proxy protocols named as HTTP Connect, HTTP Proxy, Socks, and HAProxy. 

\textbf{System-level validation:} The proxy creation and connection establishment messages generally gets logged in \verb|logcat|. We present the type of logs that are generated in \Cref{fig:logcat,fig:ps}. In both figures, the last line shows the system log.


\subsubsection{Tor-Bridge/TorPT validation}
APKs connect to Tor bridges using the Tor client program developed by the Guardian Project \cite{guardian_project}. There have been several developments in the recent past for easier integration of TorPT, into an APK. 

\textbf{Network-level validation:} At network level, it is almost impossible to detect if the (IP, port) to which the APK is connecting to is running a Tor bridge/PT. 

\textbf{System-level validation:} At system level, we can observe if the service related to a specific TorPT has started, or specific bridge libraries have been loaded. \textit{E.g.,} in \Cref{fig:logcat,fig:ps}, we show that \verb|obfs4|'s and \verb|snowflake| PT's information are visible through \verb|logcat| and \verb|ps| commands output.

In \textcircled{3}, along with dynamic validation, we also perform static analysis of APKs present in the dataset obtained from step \textcircled{2}. Next, we provide details about step \textcircled{4} that processes the the output data from both static and dynamic analysis.

\subsubsection{I2P validation}
\label{sec:i2p_validation}
Every node (sender/receiver) becomes a part of the I2P network by default. Once its application starts, it acts as a router to create tunnels and send encrypted messages to peers. The router process downloads a copy of the \textit{netdb} (information of other nearby routers in the network) and starts building inbound/outbound tunnels.

\textbf{Network-level validation:} Similar to TorPT, it is not possible to detect I2P traffic using ports or addresses since it uses random high-value ports. In the absence of central databases of peers, or \textit{reseed servers}\footnote{A process that periodically gathers info from routers.}, manual peer discovery \cite{hoang2018empirical}\footnote{The I2P's official website \cite{i2p_metric} uses such measures to determine peers, but for ethical reasons anonymizes their IPs.}, a non-trivial task, is the only resort.

\textbf{System-level validation:} Once the I2P router starts and reseeding begins, we may observe that one (or more)
\textit{RouterInfo} files (\hash{router-<random string>.dat}) are downloaded.
These files have peers' IP-port pairs, some of which the router connects to, identifiable from network traffic. Another log file, generally named \textit{Eventlog.txt}, can be inspected for reseed attempts.

\subsection{Characteristics of Flagged APKs}
\label{sec:measurement_flagged_apks}

\begin{figure}[t]
	\centering
    \includegraphics[width=0.45\textwidth]{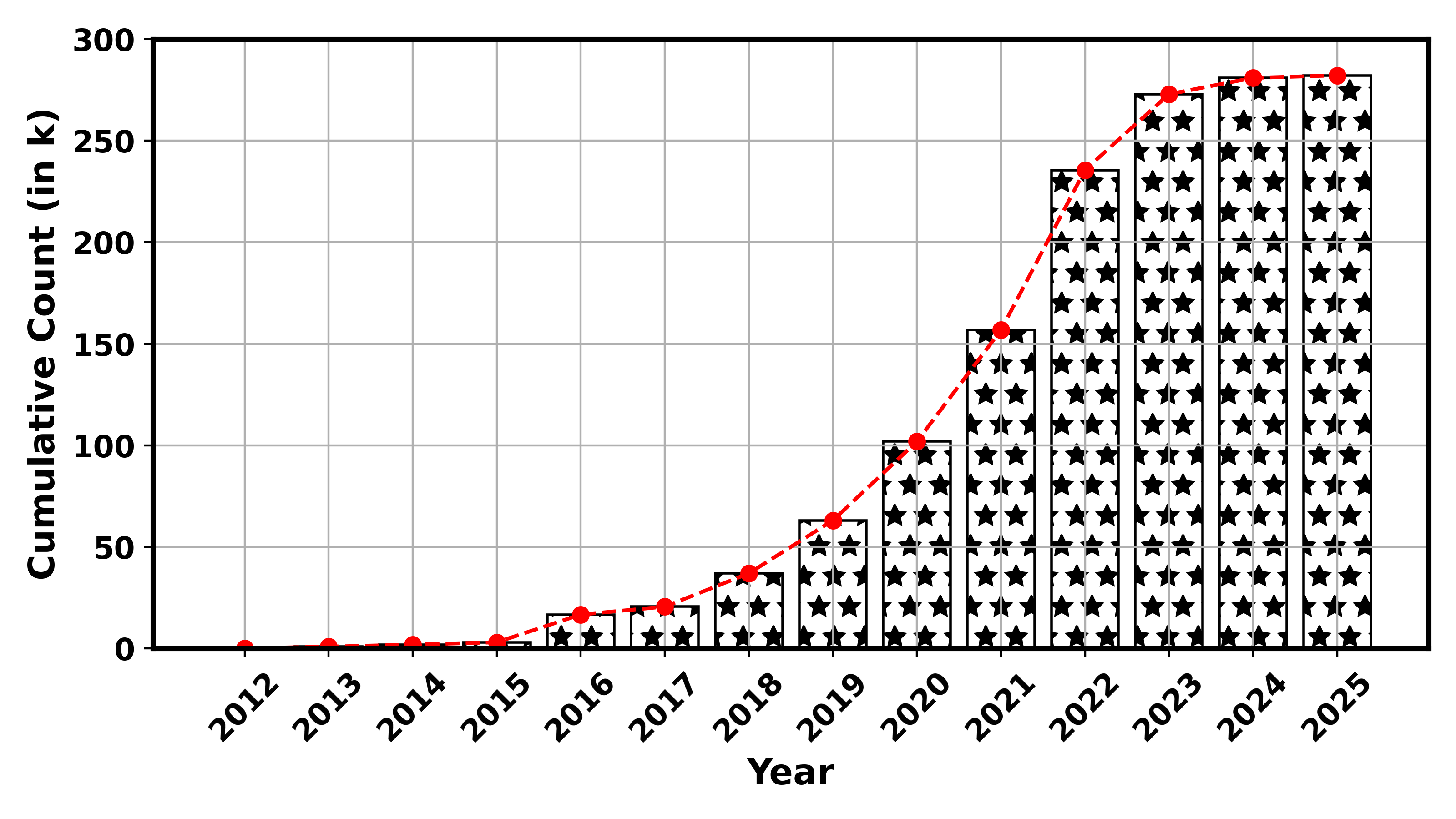}
	\caption{Yearwise cumulative count of statically validated malware APKs.}
	\label{fig:cumulative}
\end{figure}

\begin{figure}[t]
	\centering
	\includegraphics[width=0.48\textwidth]{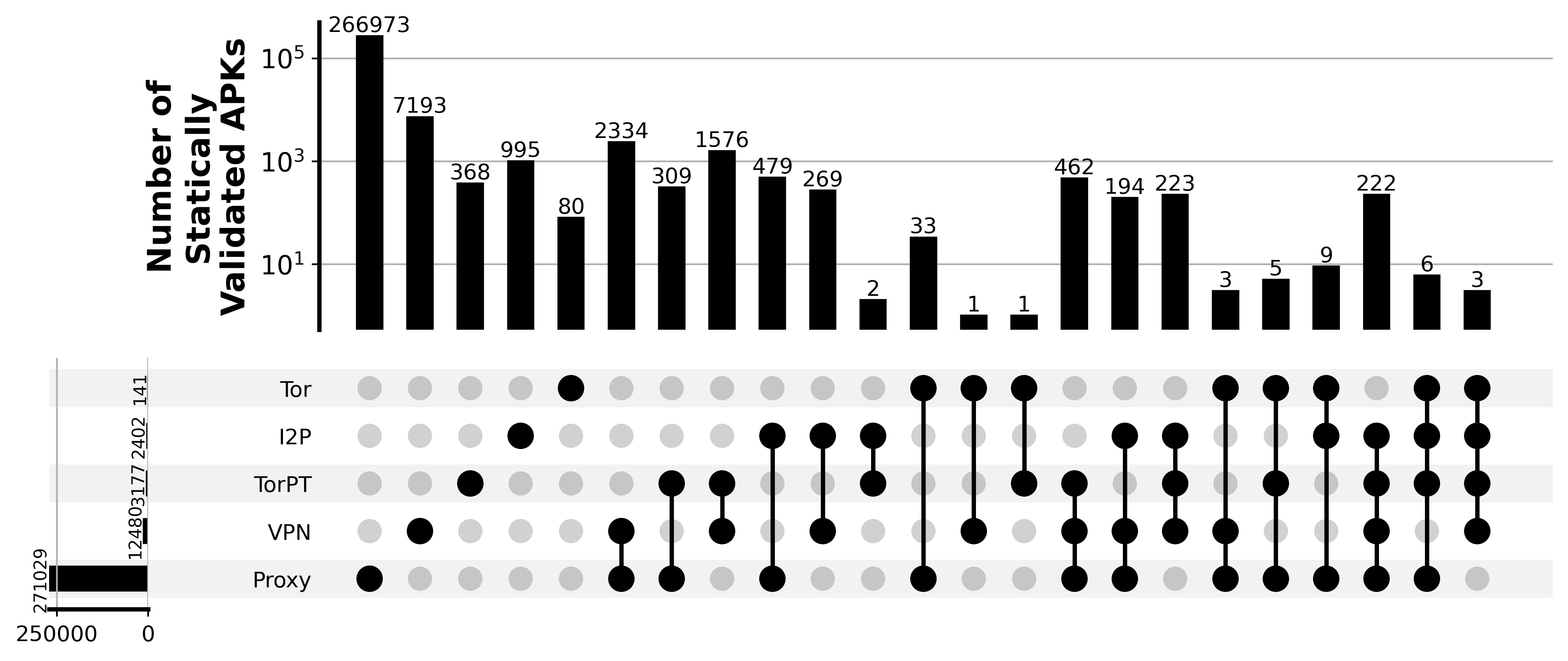}
	\caption{Distribution of APKs across individual CCs and co-occurrence of multiple CCs.}
	\label{fig:upset}
\end{figure}


     

\begin{figure*}[t]
\centering
    \begin{subfigure}{.20\textwidth}
        \includegraphics[scale=0.28]{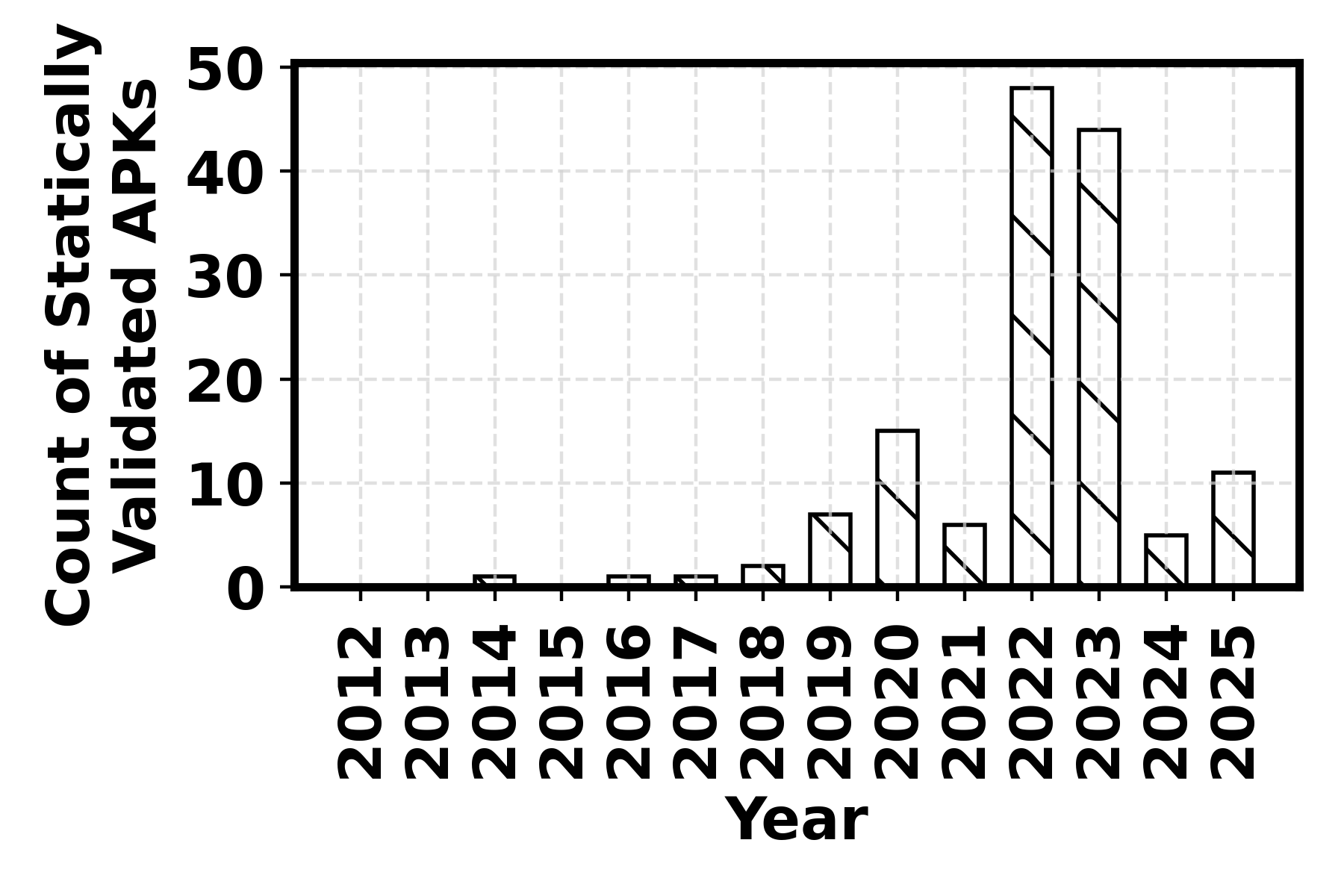}
        \caption{Tor}
        \label{fig:tor}
    \end{subfigure} 
    \begin{subfigure}{.19\textwidth}
        \includegraphics[scale=0.28]{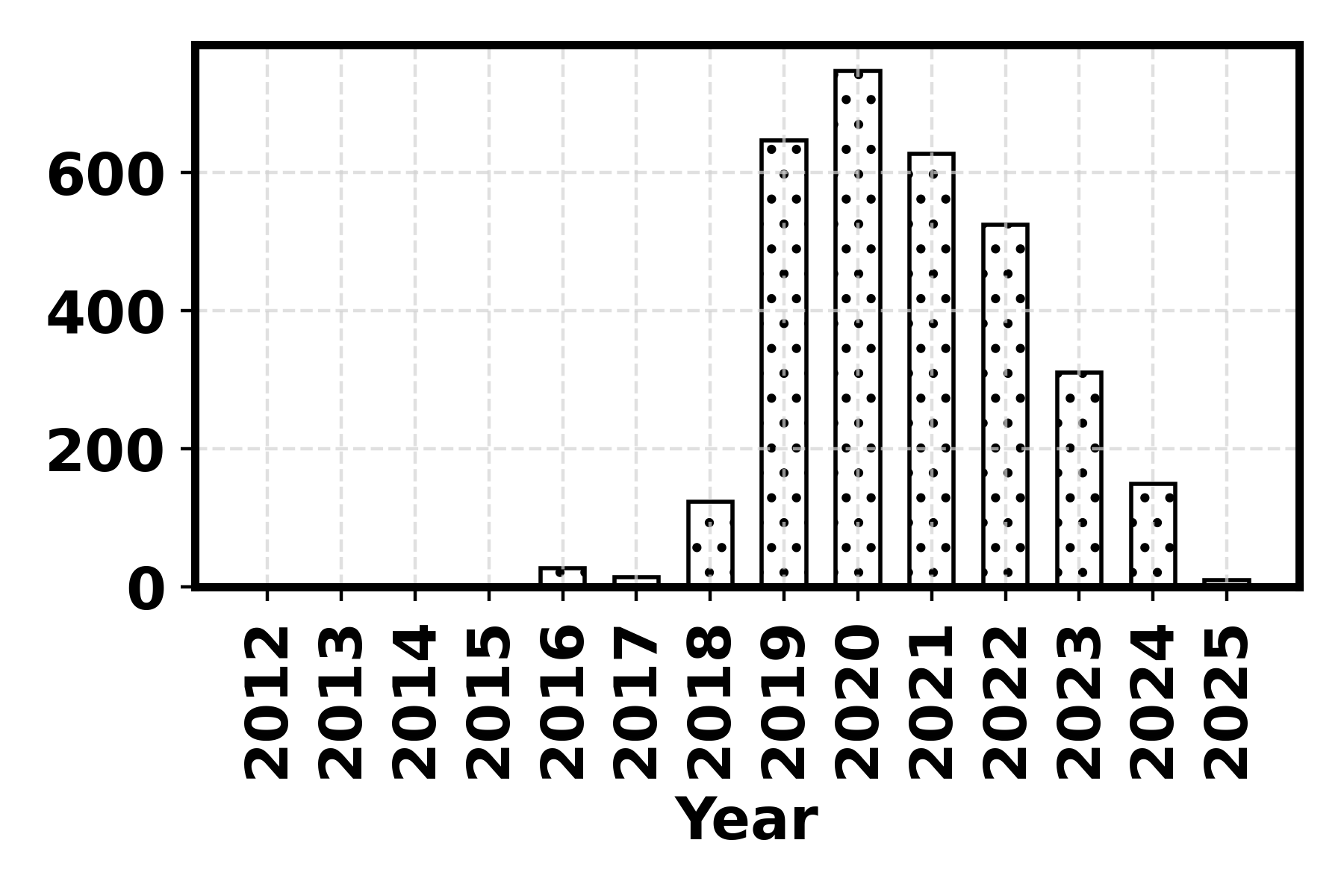}
        \caption{TorPT}
        \label{fig:torPT}
     \end{subfigure}
     \begin{subfigure}{.19\textwidth}
        \includegraphics[scale=0.28]{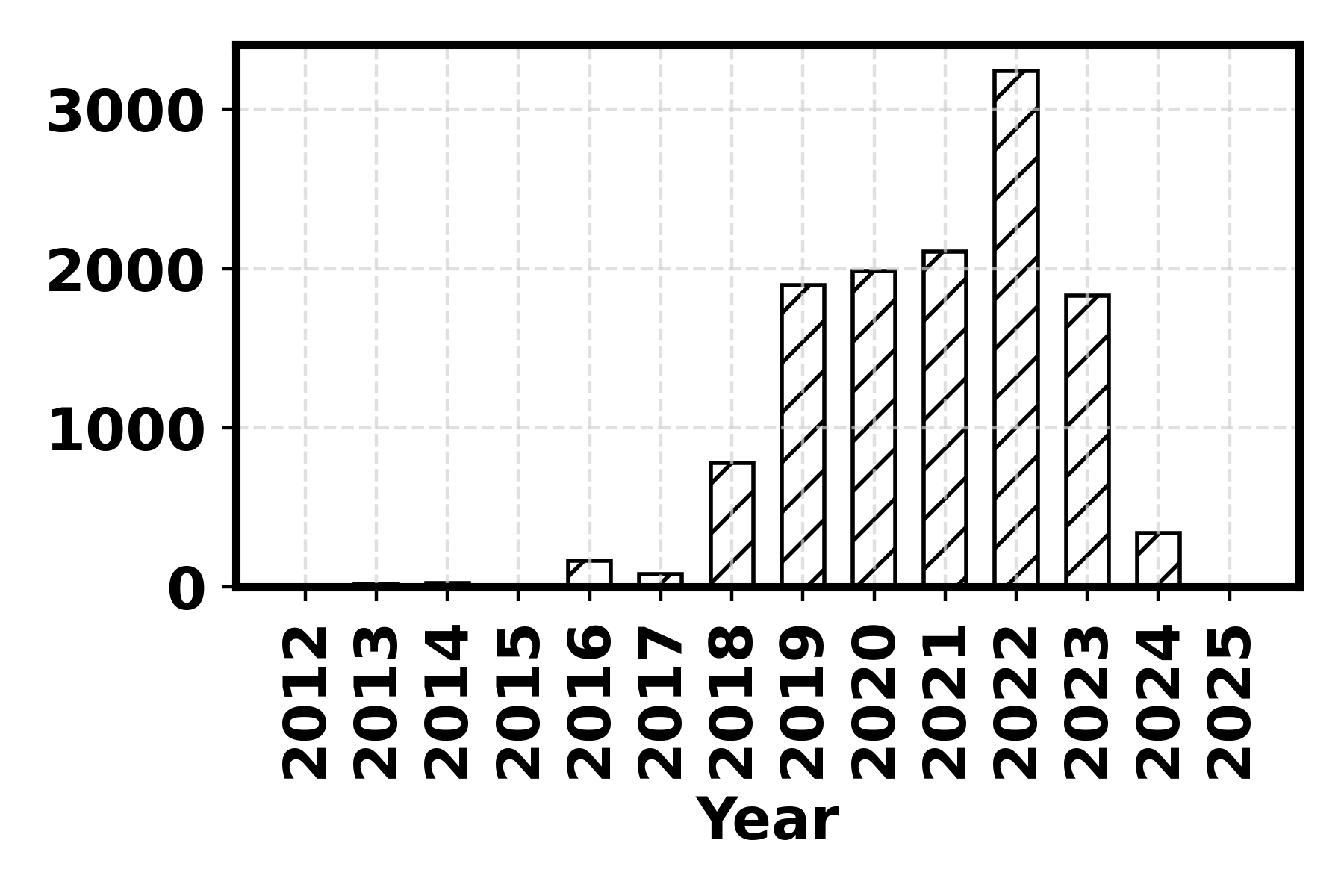}
        \caption{VPN}
        \label{fig:vpn}
     \end{subfigure}
     \begin{subfigure}{.19\textwidth}
        \includegraphics[scale=0.28]{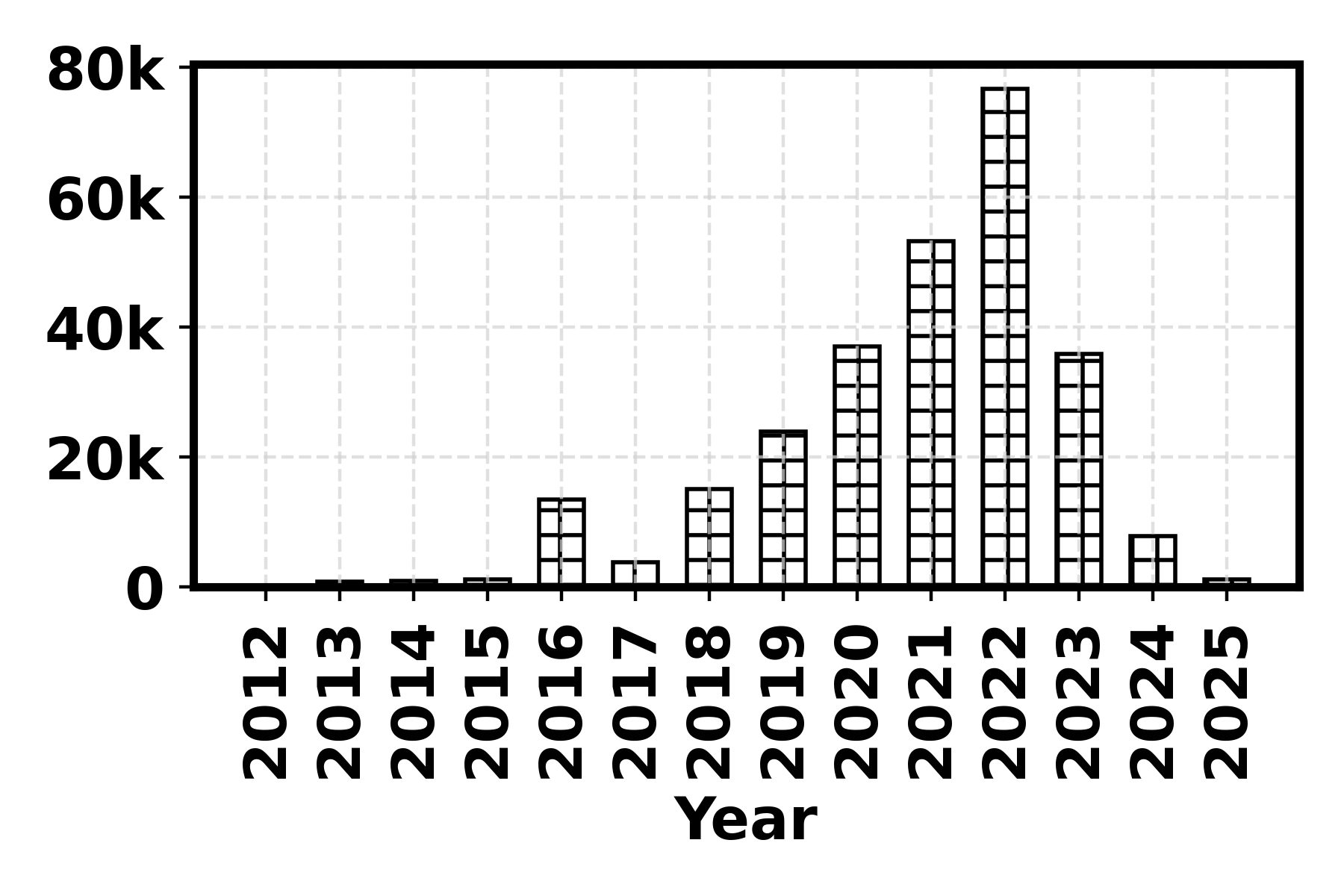}
        \caption{Proxy}
        \label{fig:proxy}
     \end{subfigure}
     \begin{subfigure}{.19\textwidth}
        \includegraphics[scale=0.28]{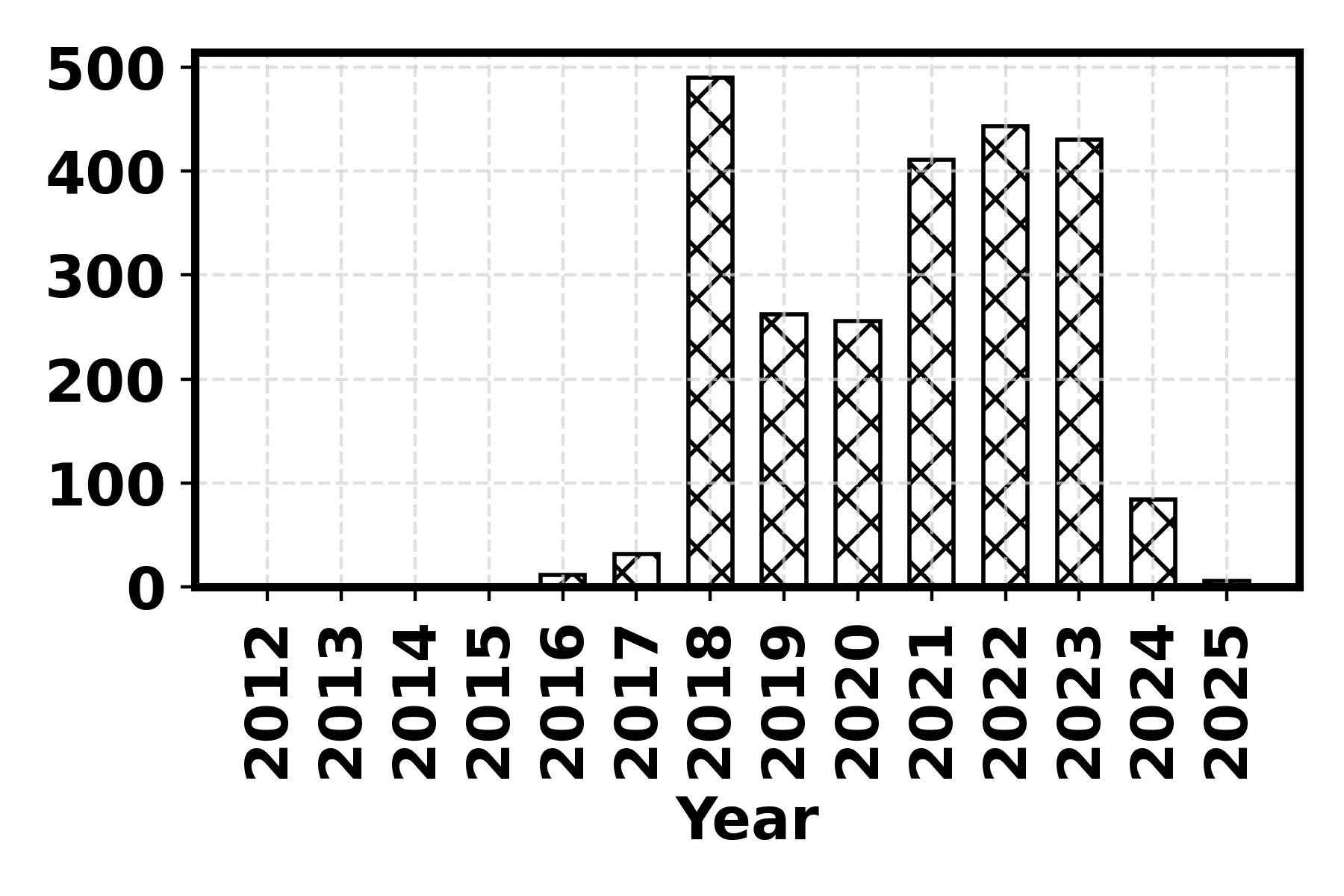}
        \caption{I2P}
        \label{fig:i2p}
     \end{subfigure}
     
    \caption{Yearwise count of malware APKs in each covert channel.}
    \label{fig:cc_yearwise}
\end{figure*}

After applying SVR to 3.6M malware APKs (\textcircled{2} in \Cref{fig:pipeline}), we obtain final dataset of malware APKs that integrate CC. To the best of our knowledge, this is the first such dataset. We considered APKs between 2009 and 2025, \textit{i.e.,} from the inception of Android OS until present. 

We perform deeper code analysis on the obtained APKs to understand their behavioral characteristics. Along with deeper code level static analysis data, we also combine dynamic data, to further understand the dynamic behavior of a portion of APKs. Thus we use both static as well as dynamic data to find the properties of the flagged APKs (for results, see \Cref{sec:sta_val_apks,sec:dyn_analisis}).


Overall, one could monitor the trends of such APKs, over the years and prominent families using such CCs. We performed deep code analysis of a few APKs from each category (\textit{i.e.,} Tor, TorPT, I2P, VPN, and proxy) in an attempt to understand the usage of CCs by malware creators. We also observed the evolution in the usage of CCs by looking at the presence of their relevant code simultaneously in a single APK. Using dynamic analysis data, we observed network level characteristics of these malware such as where they are connecting or what ports they are using \textit{etc.}. 

%% file: sections/measurement_results.tex
\section{Measurement Results}
\label{sec:measurement_results}

We present the results of our measurement study in this section, starting with the inaccuracies observed in keyword filtering, the dataset obtained after applying SVRs, and their detailed static and dynamic analysis.

\input{sections/dataset}
\subsection{CC Malware Dataset}
\label{sec:final_dataset}

After applying the SVRs on 3.6M APKs from 2009-2025, we curated our final dataset of 288k APKs with 150, 3532, 2474, 17724, and 271762 samples, using Tor, TorPT, I2P, VPN, and proxies, respectively.
We present their yearwise cumulative counts in \Cref{fig:cumulative}\footnote{For package level statistics, see \Cref{sec:package_level_stats}.}. As evident, in 2016 there were under 25k CC APKs which dramatically increased to 288k by 2025. It shows that a lot of malware APKs have the presence of CC in their code (\textit{i.e.,} 7.88\% of malware present in AndroZoo (3.6M)). 
Overall, we observed that the percentage of APKs using CC increased from 0.30\% (in 2012) to 50\% (in 2025)---a rise of 99.4\%, indicating that CC usage is growing at an alarming rate.

Our dataset reveals a complex behavior among malware creators, who often embed multiple CCs within a single APK. \Cref{fig:upset} illustrates this phenomenon.
The leftmost bar highlights the count of APKs to be 267k that only use proxies. The sixth bar (from left), represents 2.4k APKs that include both VPNs and proxies for covert communication. 
Our final dataset has 6.7k APKs with 23 CC combinations (out of 32 possible ones), showing the technical sophistication of malware creators.

\Cref{fig:cc_yearwise,fig:upset} collectively shows that proxy usage is highest individually as well as in combination with other CCs.
As evident from the yearwise statistics in \Cref{fig:cc_yearwise}, most CC APKs use proxies, with Tor being least frequently used (under 200). There seems to be a common pattern of increasing CC APK counts from 2018. The reason could be attributed to the increased malware curation and identification by AndroZoo, 2018 onwards and the dip after 2022 can be attributed to the low number of available malware in years 2023/24/25. (see \Cref{table:initial_dataset}).   


\subsection{Static Study on Our Dataset of CC APKs}
\label{sec:sta_val_apks}

\begin{figure}[t]
	\centering
	\includegraphics[width=0.42\textwidth]{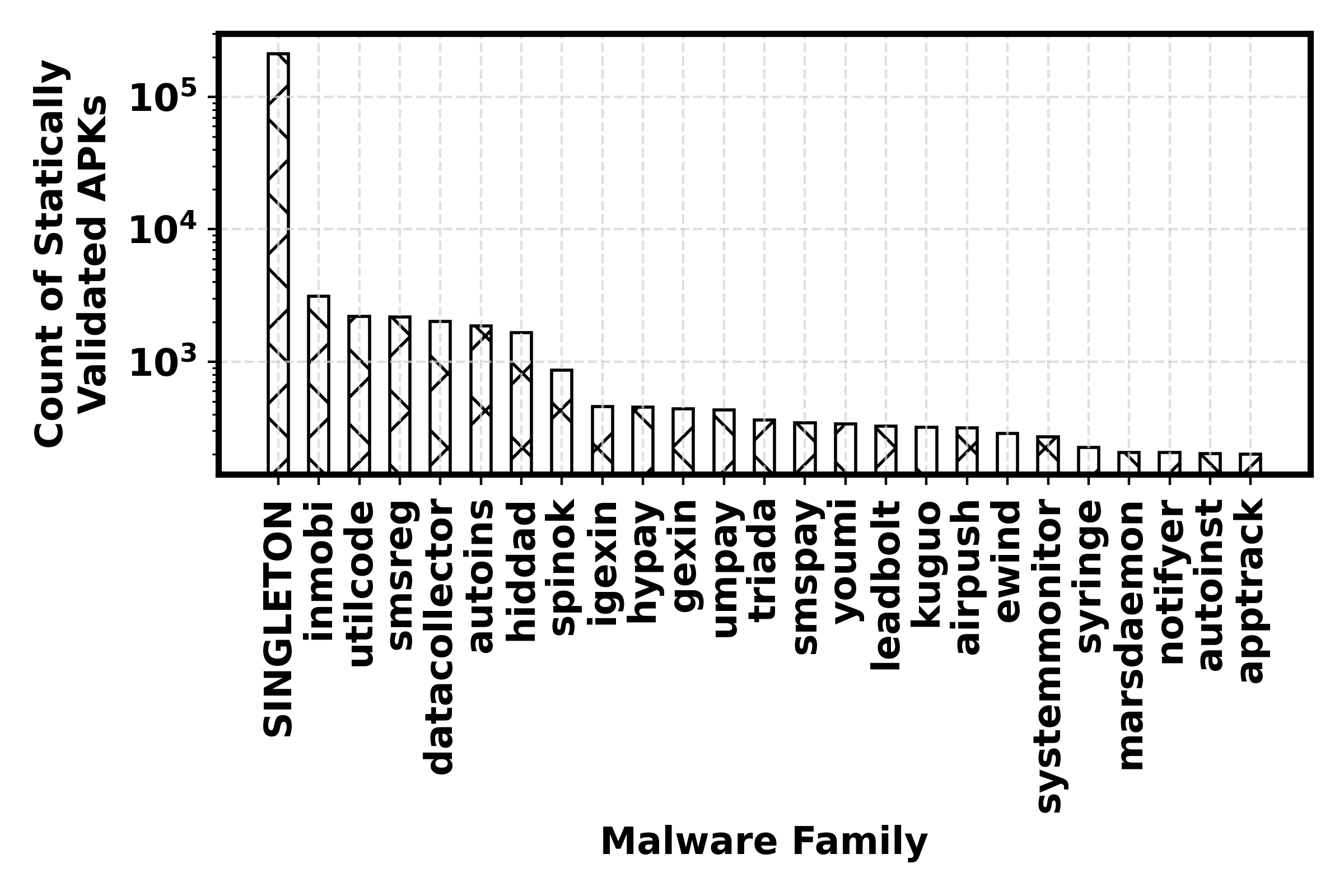}
	\caption{Top-25 malware families in our final dataset.}
	\label{fig:fam_cc_counts}
    \vspace{-4mm}
\end{figure}

\begin{figure}[t]
	\centering
	\includegraphics[width=0.35\textwidth]{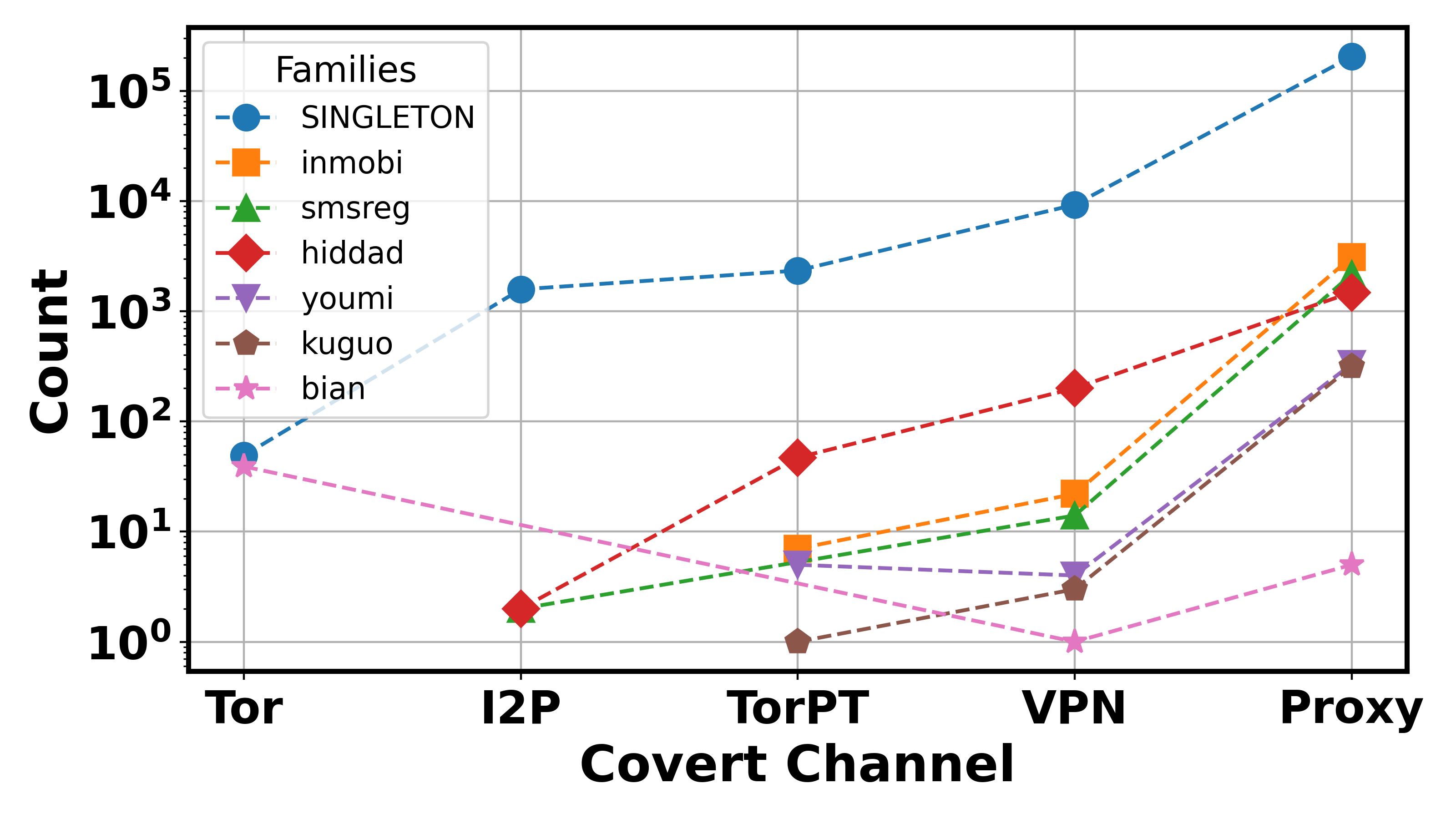}
	\caption{Malware families using multiple CCs.}
	\label{fig:fam_cc}
\end{figure}

\begin{figure*}[t]
	\centering
	\includegraphics[width=0.98\textwidth]{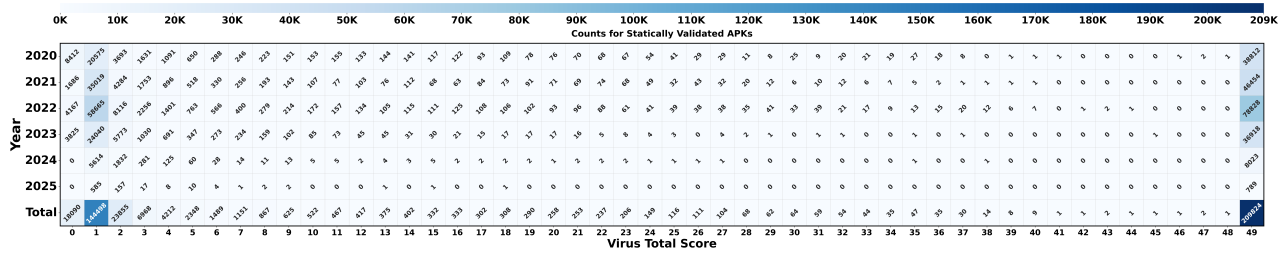}
	\caption{VT-scores and count of APKs for years from 2020-25. Heatmap for all years is shown in \Cref{sec:heatmap_all}. }
	\label{fig:heatmap_vt_scores}
\end{figure*}

To find which prominent families are involved, 
we passed all APKs' VT reports through \texttt{AVClass2} tool \cite{sebastian2020avclass2}.
We discovered 511 distinct malware families, covering our dataset.
More specifically, for Tor, TorPT, I2P, VPN, and proxy, we spotted 12, 39, 23, 127 and 477 unique families, respectively.
We observed that there are families that have APKs using diverse
CCs. \Cref{fig:fam_cc_counts} shows the top 25 families and highlights an interesting observation, that these are not reported in security blogs even though their counts are quite high.
We present the data of 6 families and various singletons in \Cref{fig:fam_cc} where it shows that \texttt{bian} family has different APKs that uses Tor, VPN and proxy CCs. Similarly, \texttt{smsreg} family uses I2P, VPNs and proxies, as CCs. We present the broader picture of top-25 families along their APK counts in \Cref{fig:fam_cc_heatmap}.

Apart from specific family counts, we observed that for a significant portion of the dataset, \texttt{AVClass2} labeled it as singletons, while for others their VT-score went down to zero. And for almost 32k APKs, their VT-score have reduced to zero, earlier reported to be non-zero by AndroZoo. 
This shows that the anti-virus engines' reports are inconsistent (refer to \Cref{fig:heatmap_vt_scores} for yearwise VT-scores and count of APKs). 
This is a well known phenomenon of engines since they frequently update their backends to improve accuracy and efficiency \cite{botacin2025towards,zhu2020measuring,wang2023re}. 

\subsubsection{\textbf{Extensive code resue by malware creators:}}
\label{sec:heavy_code_reuse}
We determine the 
number of unique APKs by counting the number of unique main packages. 
We observe that about 124k ($\approx$ 43\% of 288k) were unique. 
The multiple APKs with same main package names are likely different versions of the same APK.
The top few highest version APKs are \hash{com.qihoo.appstore} (2306), \hash{com.qiyi.video} (489), \hash{com.m4399.gamecenter} (468), and \hash{com.sogou.androidtool} (357).
These statistics indicate that malware creators heavily reuse their codebases.

\subsubsection{\textbf{APK certificates:}}
\label{sec:apk_certs}
Using the certificates embedded in APKs, one can track malware creators and take preemptive steps by eliminating or flagging them as potentially malicious software.
Developer certificates have information regarding owner, issuer, serial number, validity, certificate fingerprints, signature algorithms and version. 
We observed that 75 most frequently occurring certificates (by their owner information) were present in at least 100 APKs. The most frequently occurring certificate was present in 4405 APKs. 
We also observed that majority of the malware APKs mentioned ``Google'' to be their organization, which is misleading\footnote{There were 94k APKs with certificates having organization as \textit{Google Inc.}, \textit{Mountain View} and \textit{California}.}.

\subsubsection{\textbf{Permissions used by APKs:}}
We present the trend of permissions used in \Cref{fig:permissions_overall}. It includes \texttt{internet}, \hash{access\_network\_state} and \hash{access\_wifi\_state}. Individually for each CC, we observed the same permissions in abundance. Interestingly, there were 22 dangerous permissions among the 50 most frequently used permissions by malware.

\subsubsection{\textbf{Presence of \textit{.onion} and \textit{.i2p}:}}
We statically inspected 8k (in each CC) APKs to find any presence of .onion and .i2p domains in APK codes. We found 15 such APKs for .onion. In this set, two APKs had the same domain \footnote{\hash{b2nkt5doeamvdmjzfz7g42hk5vdtlnktlgzhel2bgjcc4v4jhnx2qrqd.onion}}. The other 13 APKs had implementation to visit onion sites through Tor. For I2P, we found 21 APKs with majority having strings such as \hash{killyourtv@mail.i2p}, \hash{HungryHobo@mail.i2p} \textit{etc.}.

\subsection{CC Usage Evolution Trend} 
\label{sec:evolution}
\begin{figure*}[t]
	\centering
    \includegraphics[width=1.0\textwidth]{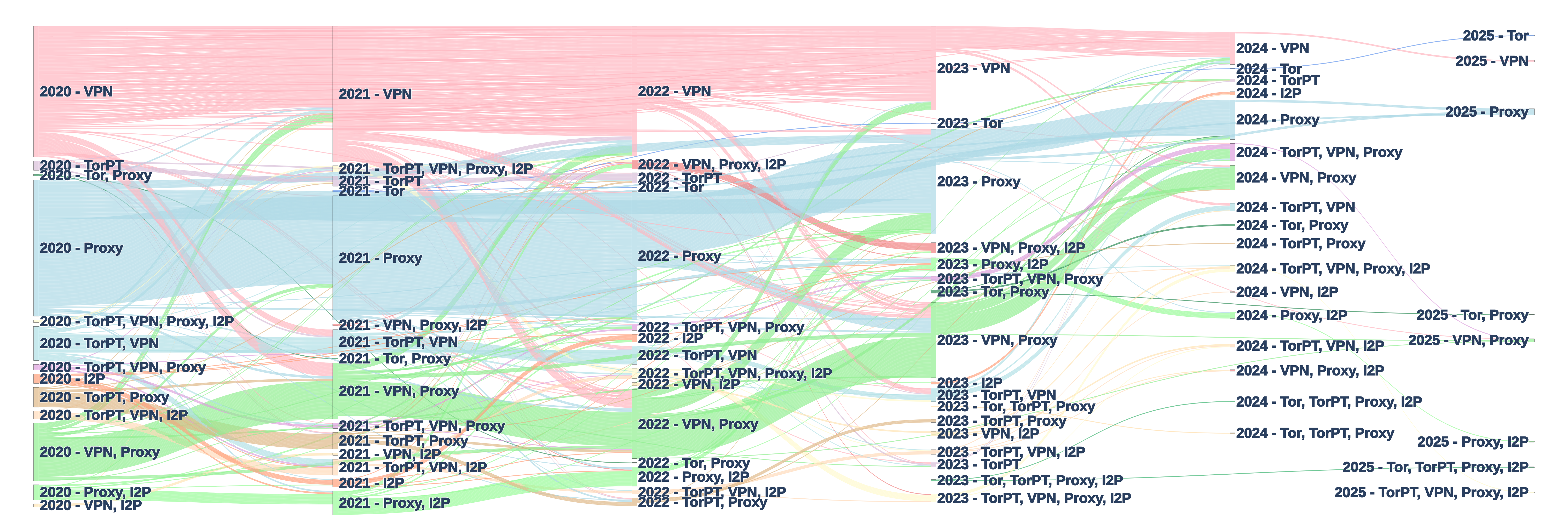}
	\caption{Evolution in the usage of CC by malware APKs from 2020 to 2025. It includes all APKs present atleast twice across the years in our dataset. The colors represent different CC combinations found in APKs.}
	\label{fig:sankey}
\end{figure*}


\begin{figure*}[t]
	\centering
    \includegraphics[width=0.95\textwidth]{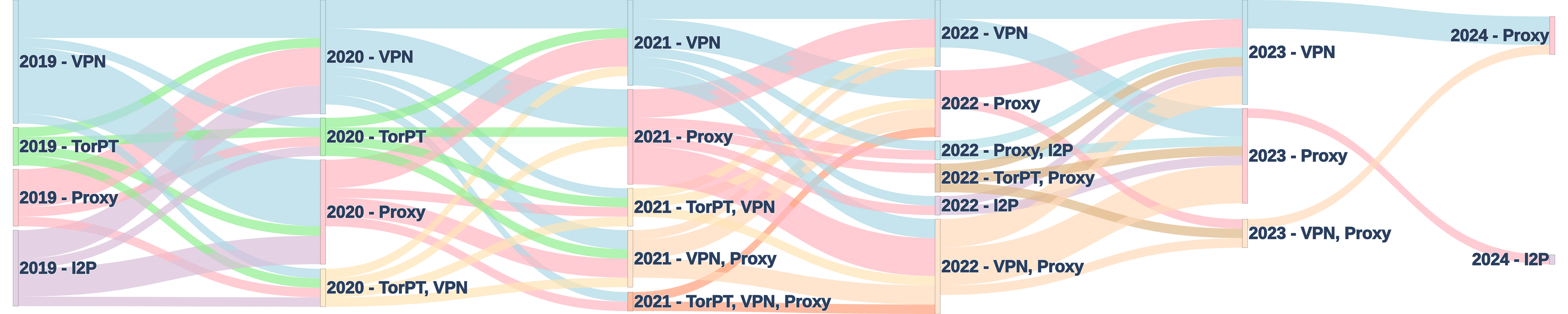}
	\caption{Evolution in the usage of 13 CC families present in all years from 2019 to 2024.}
	\label{fig:sankey_family}
\end{figure*}

We speculated that malware APKs might be evolving over time in terms of how they use CCs. To measure this, we extracted the main package (MP) names of each APK in our dataset, across years. We found that many APKs that appear in a year, have the same MPs. Thus, we selected APKs from each year and then searched for versions of the same appearing in subsequent years to study changes in CC usage. To our surprise, we found multiple APKs switching from one CC to multiple ones, and again moving back to the previous one. 
Moreover, the number of CC combinations has also been increasing over the years \textit{e.g.,} it increased from 7 to 16 combinations in a span of only three years (2017-2020) observed over complete data.
Such types of evolution can be visualized using \Cref{fig:sankey}.


For the sake of better visibility, we have only shown the evolution for six years (2020-2025) and we reduced the number of APKs in proxy CC as its in thousands. 
An example of evolution is an APK from 2021 with MP \hash{com.appatomic.vpnhub}. It earlier used VPN but later in 2022 also accommodated proxy packages. There are those APKs that earlier used both VPNs and proxies, but later switched to only using VPNs. Another interesting APK, \hash{me.talkyou.app.im}  continuously switched from using VPNs and proxies to VPNs only, and then back to the combination, and finally to proxy only, between 2020 and 2023.
We found a total of unique 378 such changes in CC preferences in our APKs. The actual number of transitions could be much higher.

\textbf{Correlation between family labeling and CC-integration:} Family labels are provided by VirusTotal and aggregated using \texttt{AVClass2} tool. 
There were 13 families (out of 511) which were consistently present in years from 2019-24. These are \texttt{autoins, smsreg, hiddad, smssend, umpay, triada, ewind, cryxos, dianjin, traca, smsspy, dataeye} and \texttt{kreditspy}. We present their yearwise CC evolution in \Cref{fig:sankey_family}.
We observed an interesting trend that CC integrations do not correlate with family labels provided by VT's anti-virus engines.
Thus, it suggests that VT engines might not be taking CC usage in consideration while making rules for family detection. 
We also observed that the top CC-integrated families appearing in 2019-2024, changed their CC channels over time, except for those that use Tor.
This potentially shows that Tor might not be as popular, compared to other CCs.




\begin{figure*}[h]
\centering
\begin{subfigure}{.32\textwidth}
  \includegraphics[scale=0.25]{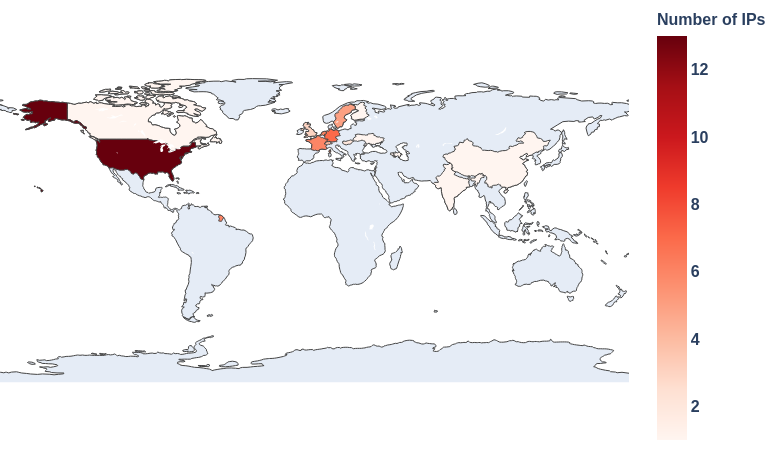}
  \caption{IPs across the globe.}  
  \label{fig:world heatmap}
\end{subfigure}
\begin{subfigure}{.32\textwidth}
  \includegraphics[scale=0.32]{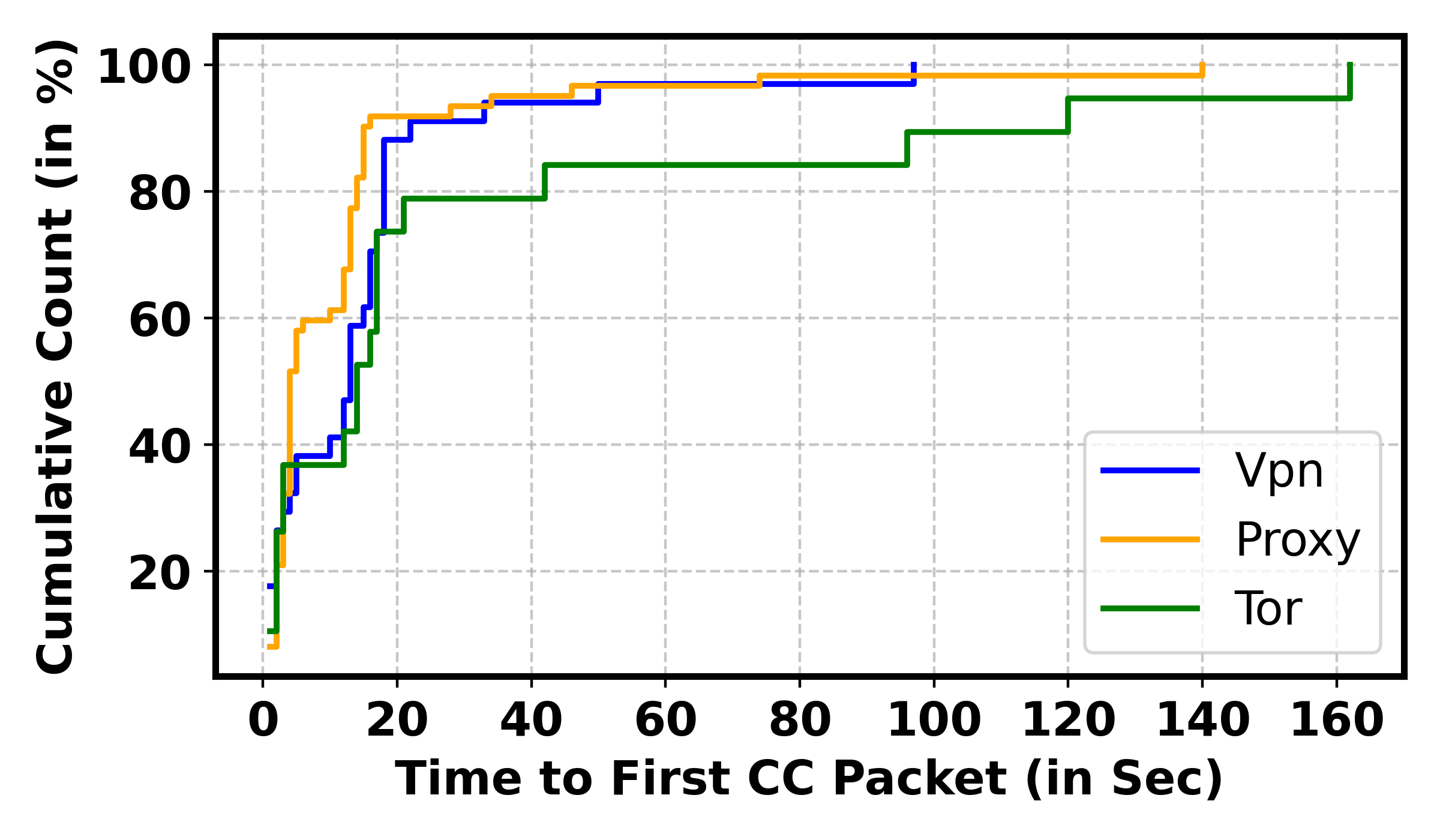}
  \caption{Time to first packet and their counts.}  
  \label{fig:ttfp}
\end{subfigure}
\begin{subfigure}{.32\textwidth}
  \includegraphics[scale=0.32]{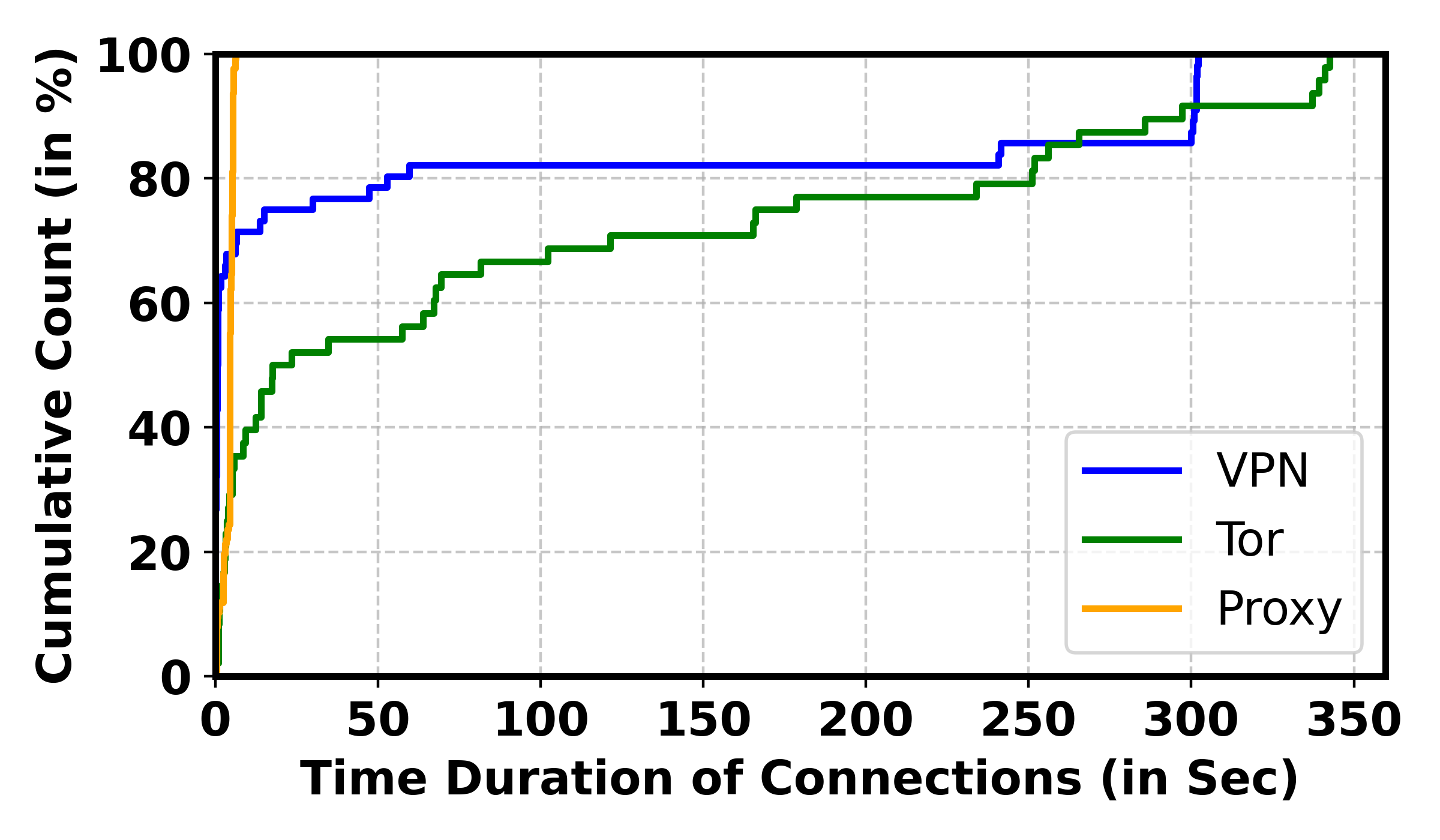}
  \caption{Connection durations and their counts.}  
  \label{fig:conn_duration}
\end{subfigure}
\caption{Dynamic analysis results with respect to validated CC connections.}
\label{fig:dyn_ip_cc}
\end{figure*}

\subsection{Dynamic validation}
\label{sec:syn_val}
Since our final statically validated dataset contained over 288k APKs, it wasn't feasible to perform dynamic validation for all the APKs. Thus, we selected APKs from each category in a manner that we covered all CC. For Tor and I2P, we selected all their flagged APKs. And for others we selected upto 100 APKs from each month from all years. Thus overall, we dynamically executed 13,697 APKs and captured \verb|pcaps| for around 6,495. Remaining APKs did not produce any traffic, likely evaded the execution environment.
We present success rate of dynamic runs over years in \Cref{fig:succes_rate_dynamic}. The dynamic execution took around 90 days to complete.
We first extracted IPs from the \verb|pcaps| and used our methods explained in \Cref{sec:pa_dynamic_validation} for each type of CC. As a result, the number of distinct validated IPs were 59\footnote{Maghsoudlou \textit{et al.} \cite{maghsoudlou2023characterizing}'s tool detected incorrect protocols (\Cref{fig:pam_results_protocols}).}. Apart from these, we observed 1014 unique IPs with unknown protocols. Upon checking the unknown protocol IPs in \hash{whatismyip.com} service, it returned 80 services with 19 proxies, 60 VPN servers and 1 Tor exit node, respectively  (see \Cref{fig:whatismyip_plot}).


We performed system-level validation steps on the dynamic data of successfully executed APKs. We observed \verb|logcat| entries for 15, 6, 10 and 16 APKs from Tor, TorPT, VPN and proxy categories. We could not find network-level evidences for these APKs except for one VPN and two Tor-based APKs. Thus, a total of 87 APKs have been validated at either system or network level, except for 3 with both type of validations.


\subsection{Dynamic Study of APKs}
\label{sec:dyn_analisis}

\subsubsection{Network level behavior}
\label{sec:dyn_study}

Malware connects to different CCs across world. To observe this, we first performed IP geolocation using MaxMind web-services \cite{maxmind}. As per Gharaibeh \textit{et al.} \cite{gharaibeh2017look} MaxMind has 95\% country-level accuracy. We present the results in \Cref{fig:world heatmap} using heatmap of validated IPs present in different countries. MaxMind could not identify the country for a few IPs, which were for APKs using VPNs. We observed maximum concentration IPs in US (13), Germany(7) and France(6). Overall we saw connections to 17 different countries
. This might happen because the C2 servers (or peer bots) are located in countries closer to the VPN end-points. Apart from the validated CC IPs, we observed that there were around 19k non-Google IPs (\textit{i.e.,} largely unrelated to Android services) where malware tried to connect, spread across 85 countries\footnote{We present the world heatmap in \Cref{fig:country_heatmap_overall_ips}}.



We also observed an interesting trend where multiple malware tried connecting to the same CC end-point IP. However, on each IP, we only found one kind of CC service running. Apart from that, we observed 42, 11 and 6 unique IPs corresponding to Tor, VPNs and proxies, respectively, to which the APKs connected. For Tor, we observed 12 unique destination ports (80, 443, 995, 6881, 8080, 9001, 9002, 9004, 9200, 9090, 30006, 60784). Tor ports such as 9200, 30006 or 60784 appear across multiple consensus files; e.g., 72 unique guards used port 9200 in Dec.'24 \cite{con_file}.
Similarly, we observed, four ports (80, 443, 1194, 8090) for VPNs and three ports (80, 8090, 50000) for proxies. 


\subsubsection{Liveness of CC servers}
\label{sec:liveness_of_cc_servers}
\Cref{fig:ttfp} shows the time to first CC packet. We can see that 80\% of connections receive their first packet in under 20 sec; for the remaining, it goes up to 160 sec.
Further, \Cref{fig:conn_duration} shows connection durations. Some Tor-based apps maintain active CC connections for the full 5 minutes, while proxy apps often terminate within seconds corroborating with the findings of Mehanna \textit{et al.} \cite{mehanna2024free}. Certain VPN apps disrupt internet connectivity during runtime. 




\subsubsection{rDNS lookups}
\label{sec:rdns_lookup}
Upon inspecting the packet types, we observed that only 42 malware performed DNS lookups before making connections.
The majority 
used Tor as CC. We also performed rDNS (revere DNS) on IPs seen in the traffic in order to find out the domain names associated with them. Overall, 3154 queries succeeded.
We observed that in Tor, the majority of domains were \hash{ip-api.com} (used for IP geo-location), while for VPNs, they mostly were for \hash{surfshark.com}, and a few for Google services. For proxies, we observed that the majority used HTTP proxies and performed DNS queries for \hash{jovotech.com} and \hash{afdvr.com}\footnote{\hash{jovotech:}It shows domain for sale. It might have been active at malware origin. \hash{afdvr}:Some agency working on surveillance.}.

%% file: sections/dataset.tex
\subsection{Static Validation over Keyword Filtering}
\label{sec:keyword_filtering}

As per our pipeline in \Cref{fig:pipeline}, we performed keyword search \textit{e.g.,} `tor', `vpn' on VirusTotal (VT) reports to come up with static validation rules for APKs already analyzed by VT. 
However, as already mentioned in \Cref{sec:step_1}, our goal was not to only rely on VT reports. Instead, extract the source code for malware APKs (to develop SVRs), especially for the ones, not analyzed by VT (\emph{e.g.,} upcoming APKs). 
Additionally, one would require VT subscription to obtain reports for thousands of malware APK hashes.
While explaining the rule formation process in \Cref{sec:step_1}, we pointed out that VT reports are not accurate in spotting APKs that use CC. 
Thus for finding out such APKs we had to perform static validation on all the malware APKs that were present in any year (with VT-score $>$ 0). It significantly increased the time to discover such APKs (45 days for the complete static validation).

After running keyword filtering (KF) and static validation (SV) of APKs, we compare their statistics (in terms of accuracy). 
We collected and analyzed 3.6M VT reports, and SV has been performed on 3.6M APKs. We compared the results of both the process -- (1) KF and (2) SV. 
We observe that KF identified around 102k CC APKs, while SV spotted 288k. 
Considering SV as a reference, KF is 20\% accurate and it has 79\% false rates. More precisely, precision and recall values are 20\% and 7\%, respectively.

%% file: sections/discussion.tex
\section{Discussion}
\label{sec:discussion}





\subsection{Lessons Learned}
\label{sec:lessons_learned}

\subsubsection{Matching at code-level}
\label{sec:matchscope}
Using \texttt{Matchscope}~\cite{feng2024accurate}, we compare two APKs, generating a summary that maps classes and methods from the first to the second. This work focuses on class-level matching, considering APKs matched if the first APK's packages, classes and methods, have a correspondence in the second APK.  Otherwise, the second is deemed to have a different implementation of the CC.

\textbf{Application over proxy-based APKs}:
Analyzing 11 proxy APKs with \texttt{MatchScope}, we identified 9 groups with similar codes. Matching these across 200k+ proxy APKs yielded 1,679 matches (\textit{i.e.,} 0.6\%, see \Cref{table:matchscope_proxy}).
The low success rate likely stems from considering only one version of each proxy APK, as malware creators reuse code but modify them to suit their needs, resulting in multiple library versions.

\begin{figure*}[t]
	\centering
	\includegraphics[width=1.0\textwidth]{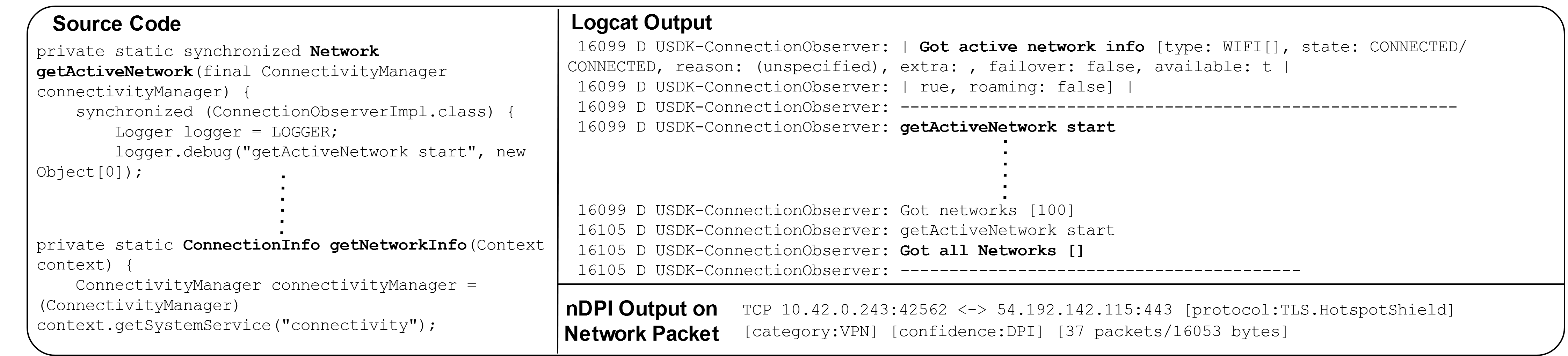}
	\caption{Source code and corresponding \texttt{logcat} output.}
	\label{fig:apk_vpn_deepcode}
\end{figure*}

\subsubsection{Obfuscation handling} 
\label{sec:obfuscation_handling}

We used \verb|LibScan| to identify CC in obfuscated APKs, by matching validated CC libraries, in invalidated APKs. For this, we considered 8053 statically validated APKs, including all type of CC APKs. Library extraction
succeeded in only half the cases (\textit{i.e.,} 3863), each bearing a CC library. Testing these 3863 libraries on 6985 APKs, we found only 39 that
appeared in 5482 APKs, alongside a highly suspicious one matching almost every APK (5106)\footnote{Its close inspection revealed that it contains code snippets for all type of CCs. And thus matching with almost every APK. Its hash is \hash{o6\_7b50b0fa883e72c339796973e2a3140a761fbae61d7036d96fab4bad35d6e50f}.}.
It shows the difficulty in handling obfuscated APKs, and the need for efforts to understand packing/obfuscations.

\subsubsection{Reality of dynamic analysis} 
\label{sec:reality_of_dynamic}
We automated APK dynamic analysis using methods similar to \textit{COMEX}\cite{umayya2024comex}, but
observing CC creation was inconsistent, as it totally depends on when the malware would want to create CC. Running malware for extended periods raises ethical concerns. Thus the absence of evidence doesn't always rule out CC creation at later stages. Future efforts should focus on artificially trigger CC creation, on the lines of Davanian \textit{et al.} \cite{davanian2024c2miner}\footnote{Although their case was for IoT malware where they clearly mentioned that it won't work for malware using CCs.}.  

\subsection{Use Case: APK Behavior Analysis}
\label{sec:apk_behav_analysis}

We selected a few APKs and observed their runtime behavior to spot CC creation and network communication.

\noindent \textbf{Case-I VPN APK=\hash{com.junkcleanner.cleanmaster.supercleaner:}} It was flagged with VPN SVR. Upon installation and execution, and on running \verb|ps|, we observed an entry named \texttt{do\_epoll\_wait} along with the APK's main package name. Further, the application 
sought permission to establish \textit{global-level} VPNs, but eventually it used only \textit{app-level} VPN connection through a new process. 
Every second the app collected information about nearby wifi networks and sent it over the VPN to a remote server. Further, on the collected pcap, when we applied nDPI, it confirmed the VPN connection establishment. We present the relevant source code, \verb|logcat| and \verb|nDPI| output in \Cref{fig:apk_vpn_deepcode}.

\input{sections/sota_table}

\noindent \textbf{Case-II Proxy APK=\hash{com.jovision.kowasmart}:}
Similar to VPN APKs, upon installation and execution, and on running \texttt{ps} command, we observed a similar
entry against \hash{do\_epoll\_wait}. Upon code inspection, we discovered a Java class \hash{StatConstants} with static strings of URLs to a website like \hash{qq.com} and its sub-domains. \verb|Logcat| shows connection attempts, by some obfuscated class of \hash{com.tencent.stat}, to a local process running at 0.0.0.1. Multiple attempts seemed to fail reporting `socket is in use' log information. At the same time, we also observed in \verb|pcap|, an HTTP proxy connection with \hash{172.233.148.27}. The number of connections were above 40 and in each connection it showed a transfer of 10-12 packets with a few KBs of data.

\noindent \textbf{Case-III Tor APK=\hash{com.uh88ctwr.og2w6oc}:}We observed that all APKs relying on Tor, used obfuscation techniques to hide their source code. For one APK (hash=\hash{533ef52c0f797cf7e2933860034cbd8c4c2cab040c11cb3c55a599c96e519e67}), we could only observe the creation of Tor circuits through \verb|logcat| logs. For example, we discovered \hash{I OnionProxyManager: Starting Tor} and \hash{I *oxyManagerEventHandler: message: severity: NOTICE, msg: Bootstrapped 75\%: Loading relay descriptors} messages in \verb|logcat|.

\noindent \textbf{Case-IV I2P APK=\hash{pan.alexander.tordnscrypt.stable}:}
Similar to Tor, for I2P we could only verify the execution at system level. 
It has integrated one I2P binary named as \hash{libi2pd.so}. Upon installation and execution, it loads the binary and starts the I2P router as observed in the output of \hash{ps} and \hash{logcat} commands. These logs look like \hash{u0\_a279 S libi2pd.so} and  \hash{pan.alexander.TPDCLogs: readTextFileSynchronous take /data/user/0/pan.alexander.tordnscrypt.stable/app\_data/i2pd/i2pd.conf success}.

\subsection{Recreation of SOTA}
\label{sec:mal_sota_detection}

We used two state-of-the-art (SOTA) malicious traffic detectors over collected pcaps to 
spot malicious (covert) traffic
generated by malware.
We recreated Barradas \textit{et al.}'s \cite{barradas2024extending} (SOTA-I) and Qing \textit{et al.} \cite{qing2023low}'s work (SOTA-II). We used the source code and data to train the models mentioned in their paper (since the training data is not dependent on the testing malware APKs). In SOTA-I, we get four models named as C-behav-rf, C-behav-dt, C-cert-rf, C-cert-dt. In SOTA-II, we obtain a single model after training. To test these models, we selected 15 benign APKs (for each CC) and 28 malware APKs (7 of each CC category). We collected network traffic of all the APKs by executing them on a mobile phone for 5 minutes. Except, in case of the benign APKs, we ran them and accessed \textit{Tranco} \cite{pochat2018tranco} top-20 websites using them. 
We extracted features as per the SOTA's for each pcap and tested those against the respective models. Results are presented in \Cref{table:sota_results}. Each row has the results (FPR/FNR) for the corresponding category tested for both benign/malware flows against each SOTA.

Each SOTA has its specific flow and feature extraction methods, and thus the number of extracted flows, in each case, is different (see Flows columns). We followed the paper implementation as is to avoid any self-introduced biases that can effect the solutions. 
For benign APKs, it shows the percentage of flows predicted as malicious \textit{i.e., FPR}. And, for malware, it shows the percentage of flows predicted as benign, \textit{i.e., FNR}. We observe that even though SOTA-I performs well on benign flows, it makes errors upto 90\% for malware flows (high false negatives). On the contrary, SOTA-II performs well on malware flows but produces false positives, upto 98\%, which can be detrimental for benign users. 
Note that our analysis results are limited to our dataset only. While we observed SOTA detector failures, precise root-cause attribution is non-trivial because these approaches are black boxes, requiring extensive feature analysis and XAI-based explainability studies.
Thus, we conclude that further efforts are necessary to comprehend CC traffic. 



\input{sections/related_table}

\subsection{Limitations \& Future Work}
\label{sec:limitations}

\begin{enumerate}[noitemsep,nolistsep,leftmargin=0pt, labelindent=\parindent, listparindent=\parindent, labelwidth=0pt, itemindent=!]
    \item \textbf{Limited to AndroZoo}: We considered all malware APKs with a non-zero VT-score on AndroZoo \cite{androzoo} which is cumulatively 3.6M. 
    AndroZoo itself collects APKs from 15 different sources including VirusShare \cite{virusShare} and GooglePlay. 
    Thus, the study might provide a comprehensive picture of majority of malware to date.
    
    \item \textbf{Static validation limitation}: 
    Several malware download their payload or necessary execution files at runtime. 
    Thus, static validation rules cannot discover such files/libraries which get downloaded upon APK execution. 
    Current SVRs for APKs apply only to unobfuscated/unmodified code segments. For obfuscated/modified APKs, the entire library code must be matched against known libraries. We conducted this experiment, with results detailed in  \Cref{sec:matchscope,sec:obfuscation_handling}.

    

    \item \textbf{Automation for new CC}:
    Creating static validation rules without manually analyzing APKs—\textit{e.g.,} permissions, text signatures, native library loading, or new class instantiation—is challenging. Establishing such rules requires appropriate knowledge of the signatures of new CCs, which is relatively non-trivial to automate.

    \item \textbf{Future work}: We will expand the study by including other CCs that malware uses for C2/bot communications, across other platforms. Some examples include social media and IM applications (\textit{e.g.,} signal, telegram), game servers (\textit{e.g.} \verb|Pubg|, \verb|Clash of Clans|) \textit{etc.}.
    
\end{enumerate}

%% file: sections/sota_table.tex
\begin{table*}[t] 
\centering
\footnotesize{
        \caption{Traffic detection results on state-of-the-art solutions (SOTA). 
    Here for SOTA-I, we are only presenting C-behav-rf model results. For the rest three refer to \Cref{sec:sotaI_results}.} 
    
    \label{table:sota_results}
    \begin{tabular}{p{1.2cm}| R{0.6cm}|p{2cm}|R{0.6cm}|p{2cm}| R{0.6cm}|p{2cm}|R{0.6cm}|p{2cm} }
    
        \hline
         & \multicolumn{4}{c|}{\textbf{Benign APKs}}  & \multicolumn{4}{c}{\textbf{Malware APKs}} \\ \cline{2-9}  

        \textbf{Covert} & \multicolumn{2}{c|}{\textbf{\textbf{SOTA-I} \cite{barradas2024extending}}} & \multicolumn{2}{c|}{\textbf{SOTA-II} \cite{qing2023low}} & \multicolumn{2}{c|}{\textbf{\textbf{SOTA-I} \cite{barradas2024extending}}} & \multicolumn{2}{c}{\textbf{SOTA-II} \cite{qing2023low}} \\ \cline{2-9}
        
         \textbf{Channel} & \textbf{Flows} & \textbf{FPR (\%)} &  \textbf{Flows} & \textbf{FPR (\%)} & 
         \textbf{Flows} & \textbf{FNR (\%)} & \textbf{Flows} & \textbf{FNR (\%)} \\ \hline \hline

          Tor & 62 & 12.90 & 9 & \textcolor{red}{100}  & 58 & \textcolor{red}{89.65} & 34 & 0.00 \\ \hline
          VPN & 4460 & 2.06 & 2194 & \textcolor{red}{89.65}  & 642 & \textcolor{red}{96.72} & 333 & 9.00 \\ \hline
          Proxy & 908 & 2.06 & 717 & \textcolor{red}{98.46}  & 1687 & \textcolor{red}{96.79} & 809 & 22.86 \\ \hline
          TorPT & 29 & 0.00 & 18 & \textcolor{red}{83.33}  & 568 & \textcolor{red}{97.53} & 295 & 2.03   \\ \hline
          I2P & 9 & 11.11 & 4 & \textcolor{red}{50.00}  & 2623 & \textcolor{red}{95.15} & 1265 &  22.37 \\ \hline

    \end{tabular}     }       

\end{table*}

%% file: sections/related_table.tex
\begin{table*}[t] 
\centering
\footnotesize{
    \caption{Type of Study: [\textbf{Cov}: About the usage of covert channel in malware, \textbf{Eco}: Eco-system of the covert channel, \textbf{Long. Years}: Longitudinal years of study, we mention the years if it exceeds a year.].}
        \label{table:work_table}
    \begin{tabular}{p{1.9cm}| p{0.85cm}|p{0.7cm}| p{0.5cm}|p{0.5cm}|p{1.5cm}| p{8.4cm}}
    
        \hline
        \textbf{Covert Channel} & \multicolumn{2}{c|}{\textbf{Platform}}  & \multicolumn{3}{c|}{\textbf{Type of Study}} & \textbf{Most Relevant Related Work} \\ \cline{2-6} 
        
         & \textbf{Android} & \textbf{Others} & \textbf{Cov} & \textbf{Eco} & \textbf{Long. Years} & \\ \hline \hline
               
          Tor & & \centering \yes & \centering \yes & & \centering \no & TorBot Stalker \cite{fajana2018torbot}, Li \textit{et al.} \cite{li2021identification}, Dodia \textit{et al.} \cite{dodia2022exposing} \\ \cline{2-7}
           & \centering \yes & & \centering \yes & & \centering \no & Anagnostopoulos \textit{et al.} \cite{anagnostopoulos2017botnet} \\ \cline{2-7}
          &  & \centering \yes & \centering \yes & & \centering \no & Hopper \textit{et al.} \cite{hopper2014challenges}, Casenove \textit{et al.} \cite{casenove2014botnet}, OnionBots \cite{sanatinia2015onionbots}, Kang \textit{et al.} \cite{kang2015efficient} \\ \hline
                   
         VPN & & \centering \yes &  & \centering \yes & \centering \no & Vpnalyzer \cite{ramesh2022vpnalyzer} \\ \cline{2-7}
          & & \centering \yes &  & \centering \yes & \centering \no & Khan \textit{et al.} \cite{khan2018empirical} \\ \cline{2-7}
          & \centering \yes & &  & \centering \yes & \centering \no & Zhang \textit{et al.} \cite{zhang2017oh}  \\ \cline{2-7}
          & \centering \yes & & & \centering \yes & \centering \no & Ikram \textit{et al.} \cite{ikram2016analysis}, Heijligenberg \textit{et al.} \cite{heijligenberg2022leaky} \\ \cline{2-7}
          & & \centering \yes &  & \centering \yes & \centering \no & Maghsoudlou \textit{et al.} \cite{maghsoudlou2023characterizing}, Xue \textit{et al.} \cite{xue2023bypassing} \\ \hline
         
         Proxy & & \centering \yes & \centering \yes & & \centering 2016-19 & Novo \textit{et al.} \cite{novo2020flow} \\ \cline{2-7}
          & & \centering \yes &  & \centering \yes & \centering \no & Mi \textit{et al.} \cite{mi2019resident}, Choi \textit{et al.} \cite{choi2020understanding} \\ \cline{2-7}
          &  & \centering \yes &  & \centering \yes & \centering \no & Mi \textit{et al.} \cite{mi2021your} \\ \cline{2-7}
           & & \centering \yes &  & \centering \yes & \centering \no & Mehanna \textit{et al.} \cite{mehanna2024free}, Bian \textit{et al.} \cite{bian2022shining} \\ \hline

         TorPT & & \centering \yes & & \centering \yes & \no, 2022-23 & Khattak \textit{et al.} \cite{khattak2016sok}, PTPerf \cite{umayya2023ptperf} \\ \hline

         I2P & & \centering \yes &  & \centering \yes & \centering \no & Muller \textit{et al.} \cite{muller2016analysis}, Hoang \textit{et al.} \cite{hoang2018empirical}, Yin \textit{et al.} \cite{yin2019i2p} \\ \hline \hline

         \textbf{All Above} & \centering \yes & & \centering \yes & & \centering \textbf{2009-25} & \textbf{This Work} \\ \hline
    \end{tabular}        
    } 
\end{table*}

%% file: sections/related_work.tex
\section{Related Work}
\label{sec:related_work}


There are some research works highlighting the presence of malware that use CCs. But no work has previously studied their evolution over years, and nor are there readily available tools/pipelines to discover such channels in new Android malware, thousands of which appear daily.

We divided the related works into three categories. The first includes works that identify such CCs in malware. The second includes works that study the
ecosystem of various CCs. These are not categorically related to the malicious usages. And the third one includes works that explore various state-of-the-art
traffic analysis methods to see if they can be used to spot malicious (covert) traffic generated by malware. 

\noindent \textbf{Category-I Malware using covert channel:} There are no longitudinal studies that explore the usage of proxy-based CCs such as Tor, VPN, free proxies and TorPTs by malware. \Cref{table:work_table} highlights that there is one work on Android by Anagnostopoulos \textit{et al.} \cite{anagnostopoulos2017botnet} where authors proposed novel designs of botnets using Tor hidden services. There are works for other platforms such as Windows, Linux \textit{etc.} \cite{fajana2018torbot,li2021identification,dodia2022exposing,hopper2014challenges,casenove2014botnet,sanatinia2015onionbots,kang2015efficient,novo2020flow}.  \textit{TorBot stalker} \cite{fajana2018torbot}, Li \textit{et al.} \cite{li2021identification} and Dodia \textit{et al.} \cite{dodia2022exposing} worked on developing detection methods to distinguish non-Tor from Tor traffic. Only Dodia \textit{et al.} \cite{dodia2022exposing} tried to search for real malware that uses Tor. However, they found 362 such cases for Windows, which might not represent the diversities of Tor-based malware in the wild\footnote{As per VirusTotal weekly stats \cite{vt_stats}, Windows based malware are the top occurring in the wild. (As of Jan 2025)}. 

Hopper \textit{et al.} \cite{hopper2014challenges} proposed solutions to protect Tor from botnet abuse reported in August 2013 \cite{tor_blog7}. In contrast, following the same incident, Kang \textit{et al.} \cite{kang2015efficient} and Sanatinia \textit{et al.} \cite{sanatinia2015onionbots} proposed design for better botnets that cannot be easily removed. 
Additionally, the efforts by Novo \textit{et al.} \cite{novo2020flow} also aim to identify malware traffic that uses CCs. To that end, they proposed the creation of an adversarially strong DL model. However, they did not test it on real malware traffic but only used synthetic flows emulating that of a C2 server. 



\noindent \textbf{Category-II Proxy based covert channel ecosystem:}
Several works highlight the ecosystem of the CCs themselves and their security/privacy guarantees. But most do not describe their malicious usage in malware network communications. 
Zhang \textit{et al.} \cite{zhang2017oh}, Ikram \textit{et al.} \cite{ikram2016analysis} and Heijligenberg \textit{et al.} \cite{heijligenberg2022leaky} performed security analysis of Android-based VPN APKs. 
Apart from Android, others have also performed such analyses for Windows/Linux/MacOS based covert communication tools\cite{ramesh2022vpnalyzer,khan2018empirical,maghsoudlou2023characterizing,xue2023bypassing,mi2019resident,mi2021your,choi2020understanding,mehanna2024free,bian2022shining,khattak2016sok,umayya2023ptperf,hoang2018empirical,yin2019i2p}.


\noindent \textbf{Category-III Malicious traffic detection:} This is a well-explored and complex research area. Most allied works involve some form of traffic analysis to distinguish malicious from benign flows. Here we only provide an overview of the most important state-of-the-art (SOTA) detection solutions published in top-tier venues in recent years. We explored these solutions to find out if SOTA is able to distinguish between benign and malicious covert traffic. Among others like \cite{hu2022adversarial,bergman2024recognition,Wang_2024,polatidis2023vpndroid,seraj2023mvdroid,novo2020flow,chiapponi2022badpass}, we attempted to recreate some latest SOTA on malicious traffic detection such as \cite{fu2022encrypted,qing2023low,barradas2024extending} \footnote{Since Fu \textit{et al.}'s source code was not public, we could not recreate it. We were able to recreate the rest.}. We present the results later in \Cref{sec:mal_sota_detection}. Overall, we conclude that the SOTA has high FPs and thus requires more understanding of traffic that use CCs. 

Thus, firstly there is no longitudinal study over Android malware using CCs for hiding their communications. Secondly and eventually, there is an absence of tools/pipelines for discovering such APKs. As per \Cref{table:work_table}, this is the first longitudinal study involving malware APKs that have appeared in the last 16 years (since Android's inception)
that explores how CCs have been used, along with a framework to continue the research forward.

%% file: sections/conclusion.tex
\section{Conclusion}
\label{sec:conclusion}

Android malware increasingly seem to exploit covert channels (CC) like Tor, I2P, VPNs and proxies to evade detection. Despite
widespread use, no longitudinal study has examined their evolution and prevalence. 
We present a novel pipeline for detecting CC usage in malware APKs. It begins with a static validation, combining
automation and manual efforts, followed by fully automated steps. Applied to 3.6M APKs (between 2009--2025), it identified
288k that used CC--Tor (150), TorPT (3k), I2P (2k), VPNs (17k) and proxies (271k), spread across 511 families. 

Dynamic validation and analysis, using real devices, and collecting runtime network/system data (using \verb|pcaps|, \verb|logcat|, \verb|ps|,
\verb|lsof|, and \verb|netstat|), revealed that malware connected to CCs over 59 unique IPs across 17 countries. 
APK's evolutionary study revealed that malware constantly adapt CC strategies.
Further, preliminary study reveals using SOTA mechanism to tell malicious and non-malicious traffic apart, report high FPR, underscoring the need for better solutions in the future.

%% file: sections/acknowledgment.tex
\section{Acknowledgments}
\label{sec:ack}
We thank our colleagues and anonymous reviewers for their valuable feedback and suggestions.

%% file: sections/appendix.tex
\section{Appendix}
\label{sec:appendix}

\input{sections/ethics}

\input{sections/hyphanet}

\subsection{List of Keywords}
\label{sec:keywords}
For each category there is a different list of keywords. We present all these in \Cref{table:keyword_table}.

\begin{table}[t] 
\centering
\footnotesize{
  \caption{List of keywords for each covert channel.}
      \label{table:keyword_table}
    \begin{tabular}{p{2.0cm}| p{5.4cm}}
    
        \hline
        \textbf{Covert Channel} & \textbf{Keywords} \\ \hline
         Tor & dark web, *.onion, tor, torrc, orbot, onion, hidden service, tor.apk, tor2web, Ahmia, socks proxy, Tor browser, orxify, TorServices, onion service, Orchid, OnionKit \\ \hline
         VPN & virtual private network, vpn, nord, express, proton, proxy, outline, zerotier, openvpn, rav vpn, psiphon, v2ray, nymVPN, freenet, freedom network, JonDonym, Java Anon Proxy, shadowsocks \\ \hline
         Proxy & redsocks, stunnel, cntlm, http, socks, https, proxy, Proxychains, squid, Privoxy, SOCKS5, Web, Forward, Reverse, SSH, Mitmproxy, HAProxy, Switcher, Tinyproxy, Proxifier, Tunnel, SSL \\ \hline
         TorPT & decoy routing, domain fronting, snowflake, obfs4, meek, lantern, webtunnel, conjure, lyrebird, obfs, minecruft, dnstt, TorCloak, cloak, marionette, stegotorus, camoufler, massbrowser, protozoa, stegozoa, deltashaper, facet, mailet, cloudTransport, covertcast, freewave, balboa, domain shadowing \\ \hline

         I2P & *.i2p, i2p, i2pd, NTCP, NTCP2, SSU, SSU2, eepsite, i2psnark, i2pbote, BOB, SAM, SAM V2, SAM V3, I2PTunnel, i2cp, garlic routing, clients.config, router.config, i2ptunnel.config, SusiMail, i2np, garlic message  \\ \hline
        \end{tabular}        
    }
\end{table}

\subsection{SOTA-I Results}
\label{sec:sotaI_results}

\Cref{table:sota_I_results} presents results on three models namely - C-behav-dt, C-cert-dt, and C-cert-rf.

\input{sections/sota_I_table}

\subsection{Heatmap of VT-scores}
\label{sec:heatmap_all}
\begin{figure*}[t]
	\centering
	\includegraphics[width=0.98\textwidth]{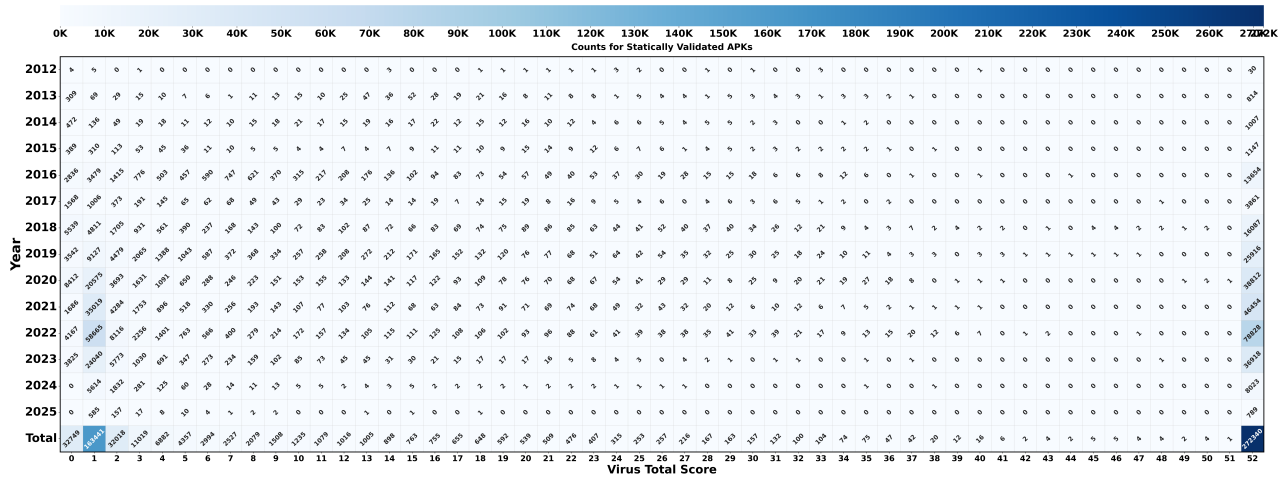}
	\caption{VT-scores and yearwise count of APKs.}
	\label{fig:heatmap_vt_scores_overall}
\end{figure*}

\Cref{fig:heatmap_vt_scores_overall} presents the yearwise count of statically validated APKs. 

\subsection{Families using multiple CCs}
\label{sec:fam_multi_cc}
\begin{figure}[t]
	\centering
	\includegraphics[width=0.47\textwidth]{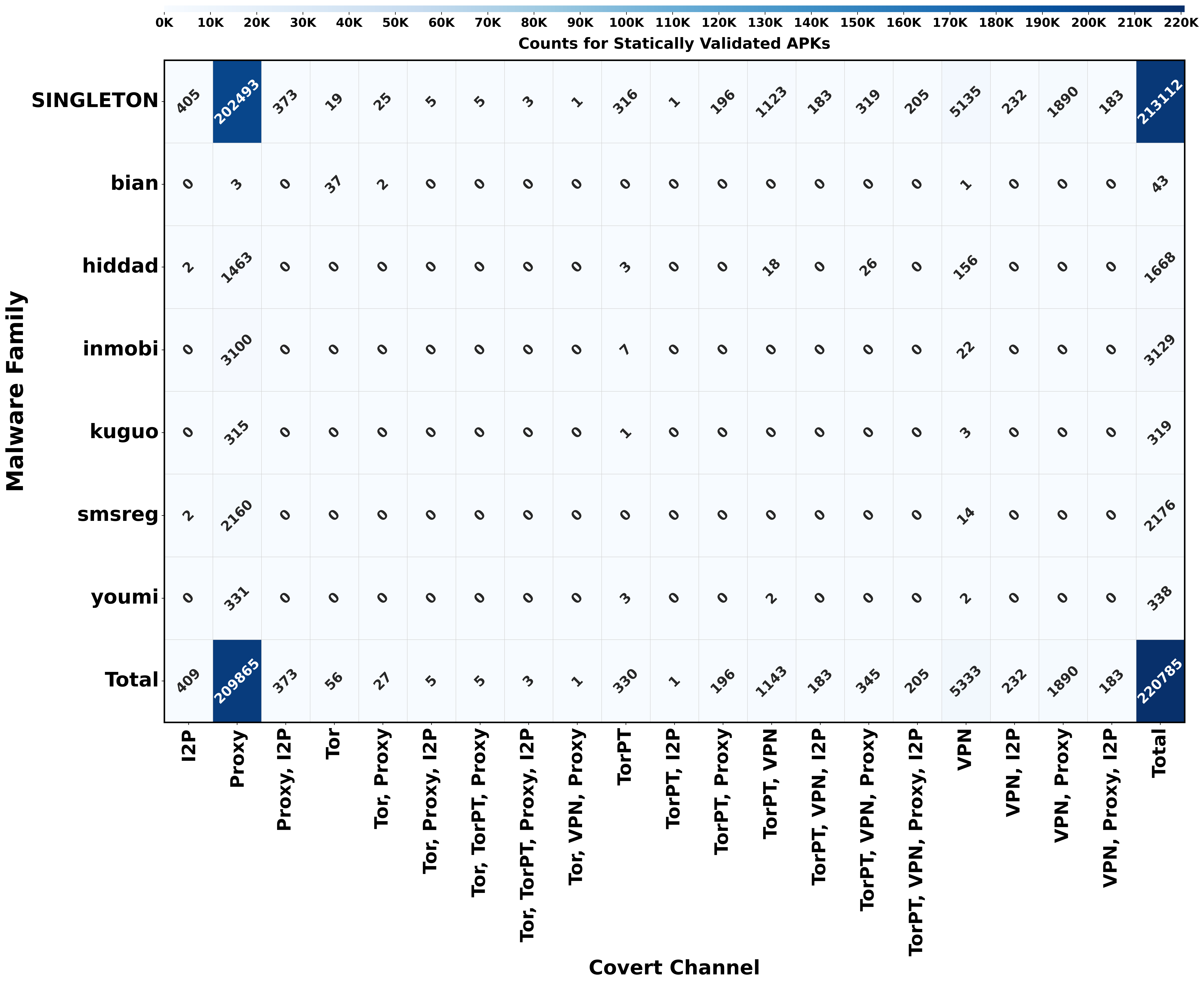}
	\caption{Heatmap of malware families using several CCs together.}
	\label{fig:fam_cc_heatmap}
\end{figure}

We present the statistics of malware families using multiple covert channels in \Cref{fig:fam_cc_heatmap}.

\subsection{Strings/Regex in Pcaps \& Results}
\label{sec:pam_paper}
\begin{figure}[t]
	\centering
	\includegraphics[width=0.47\textwidth]{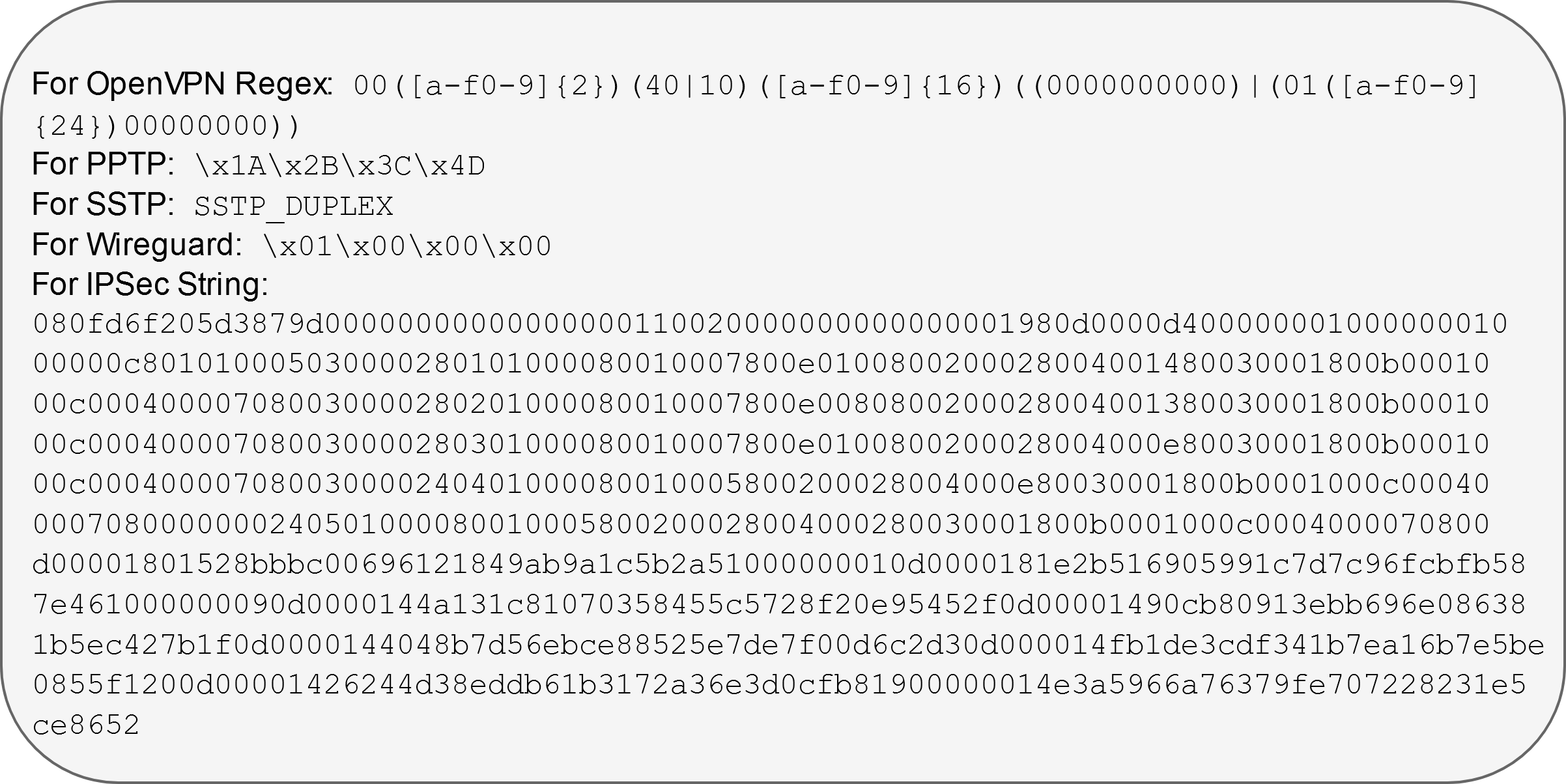}
	\caption{Strings/regex to search for in packets.}
	\label{fig:pam}
\end{figure}

We present the regex/strings in \Cref{fig:pam}. 

\begin{figure}[t]
	\centering
	\includegraphics[width=0.50\textwidth, scale=0.55]{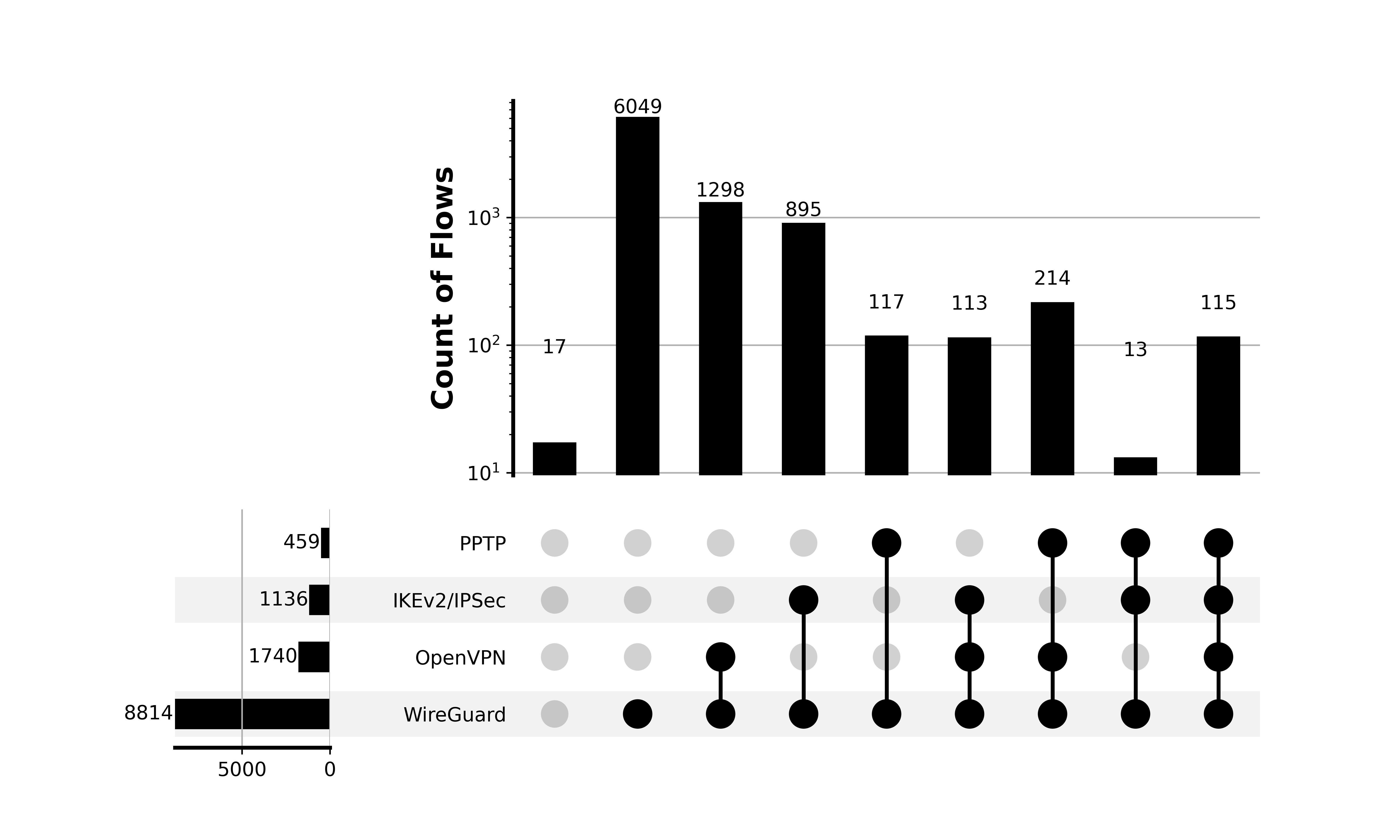}
	\caption{Detected VPN Protocols in network flows and their counts.}
	\label{fig:pam_results_protocols}
\end{figure}

\begin{figure}[t]
	\centering
	\includegraphics[ scale=0.40]{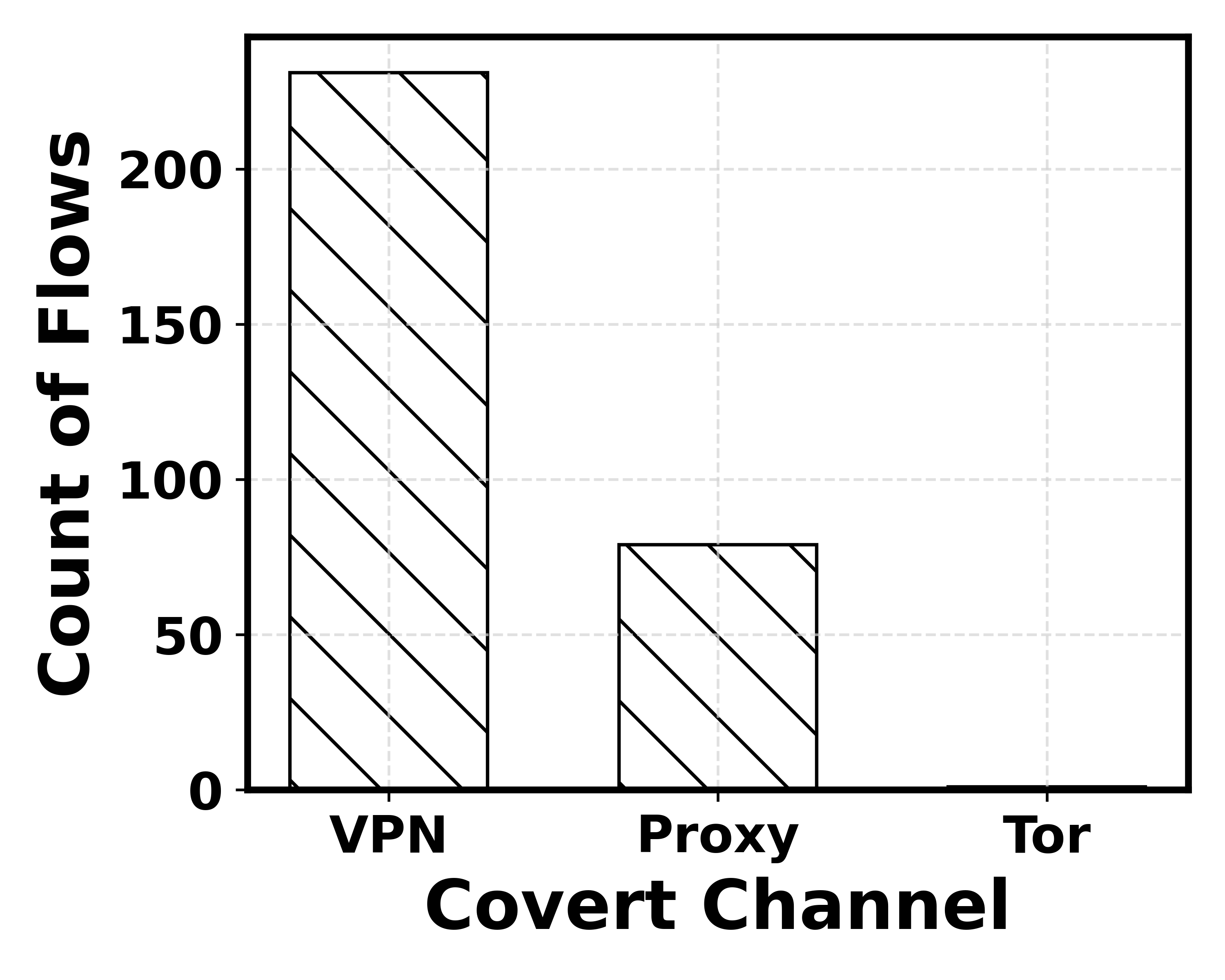}
	\caption{Presence of CC services running on (unknown protocol) IPs found via  \texttt{whatismyip.com}.}
	\label{fig:whatismyip_plot}
\end{figure}

\begin{figure}[t]
	\centering
    \includegraphics[width=0.43\textwidth]{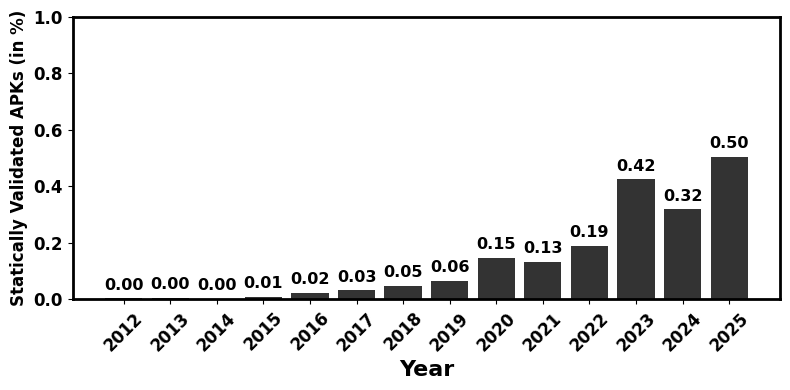}
	\caption{Yearwise percentage of statically validated malware APKs.}
	\label{fig:yearwise_percentage}
\end{figure}

\begin{figure}[t]
	\centering
	\includegraphics[width=0.43\textwidth]{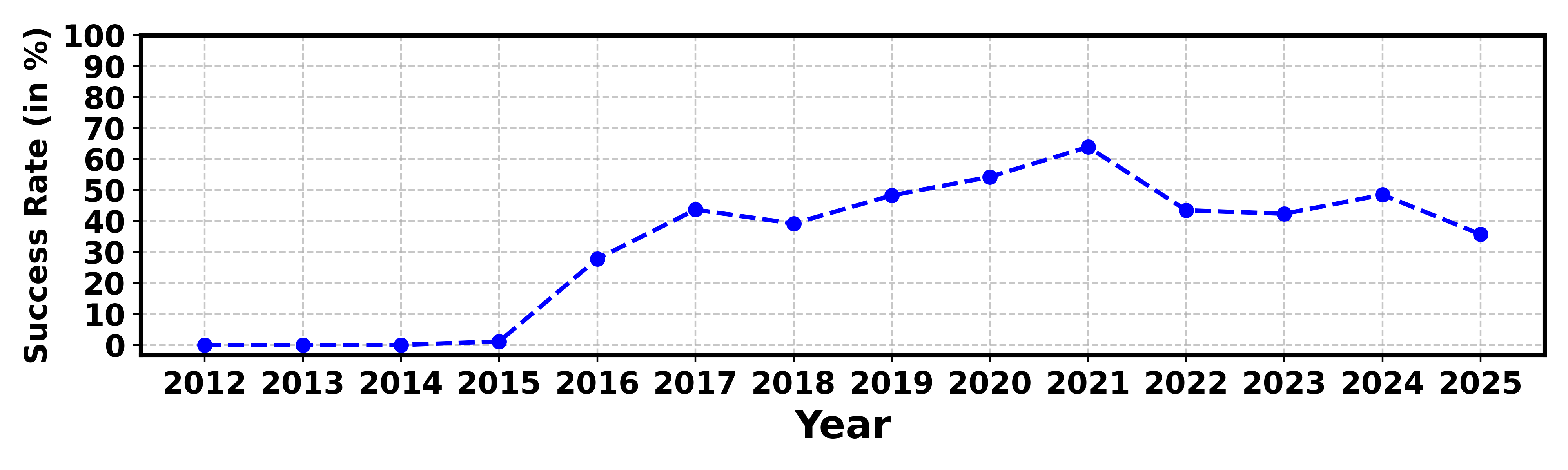}
	\caption{Yearwise success rate of dynamic analysis.}
	\label{fig:succes_rate_dynamic}
\end{figure}

\begin{figure}[t]
	\centering
	\includegraphics[width=0.36\textwidth]{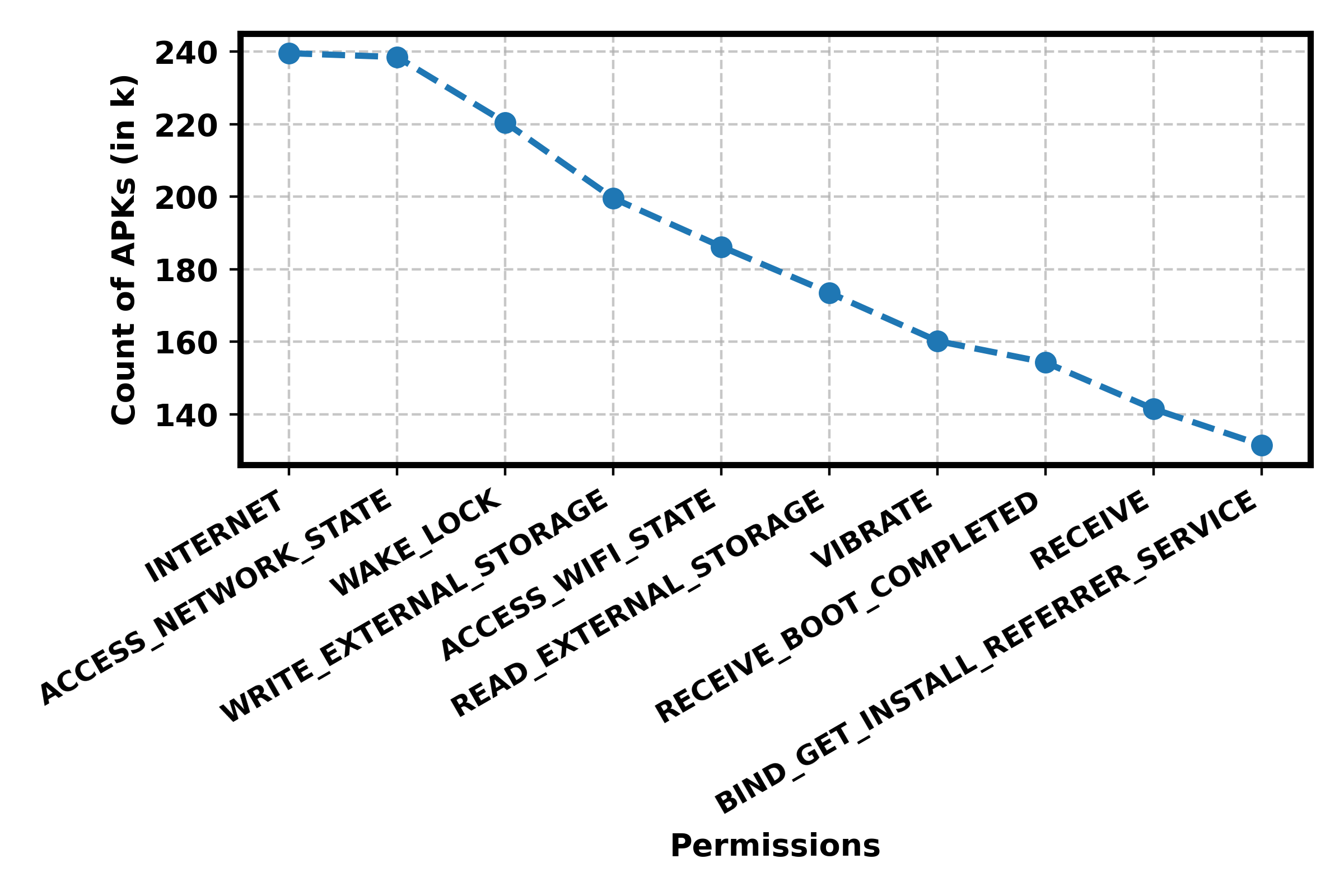}
	\caption{Top 10 permissions used by malware APKs.}
	\label{fig:permissions_overall}
\end{figure}

\begin{figure}[h]
	\centering
	\includegraphics[width=0.40\textwidth]{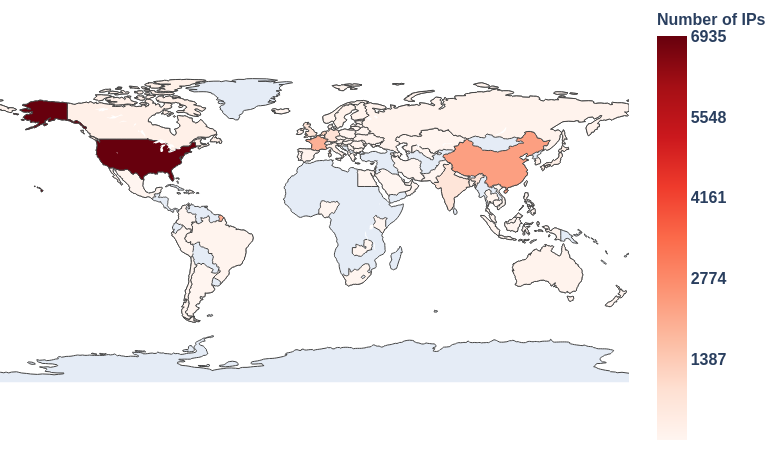}
	\caption{Malware IP connections across the globe.}
	\label{fig:country_heatmap_overall_ips}
\end{figure}

    

\begin{figure}[!t]
\centering
    \begin{subfigure}{\textwidth}
      \includegraphics[scale=0.50]{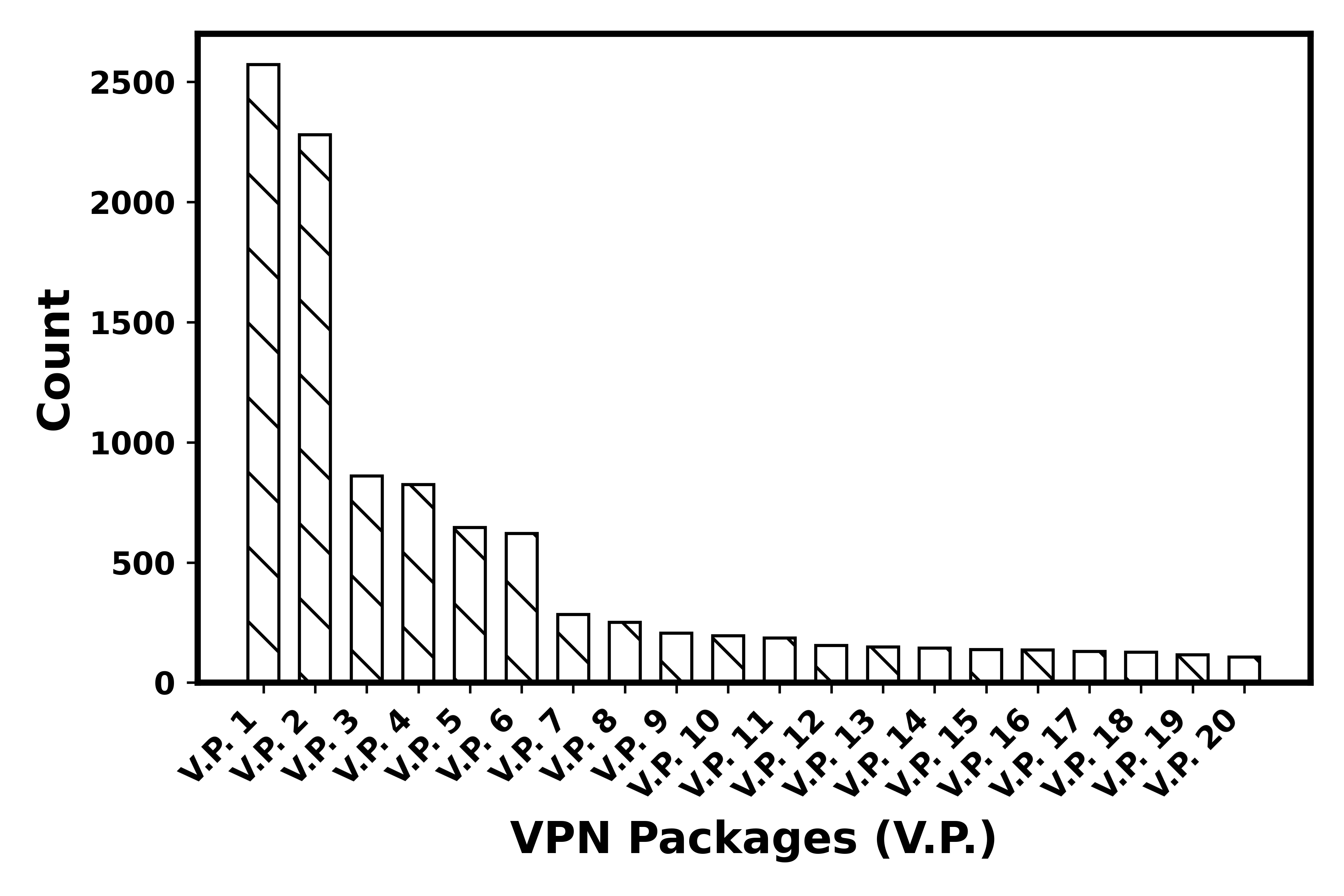}
      \label{fig:vpn_freq}
    \end{subfigure}
    \begin{subfigure}{\textwidth}
      \includegraphics[scale=0.50]{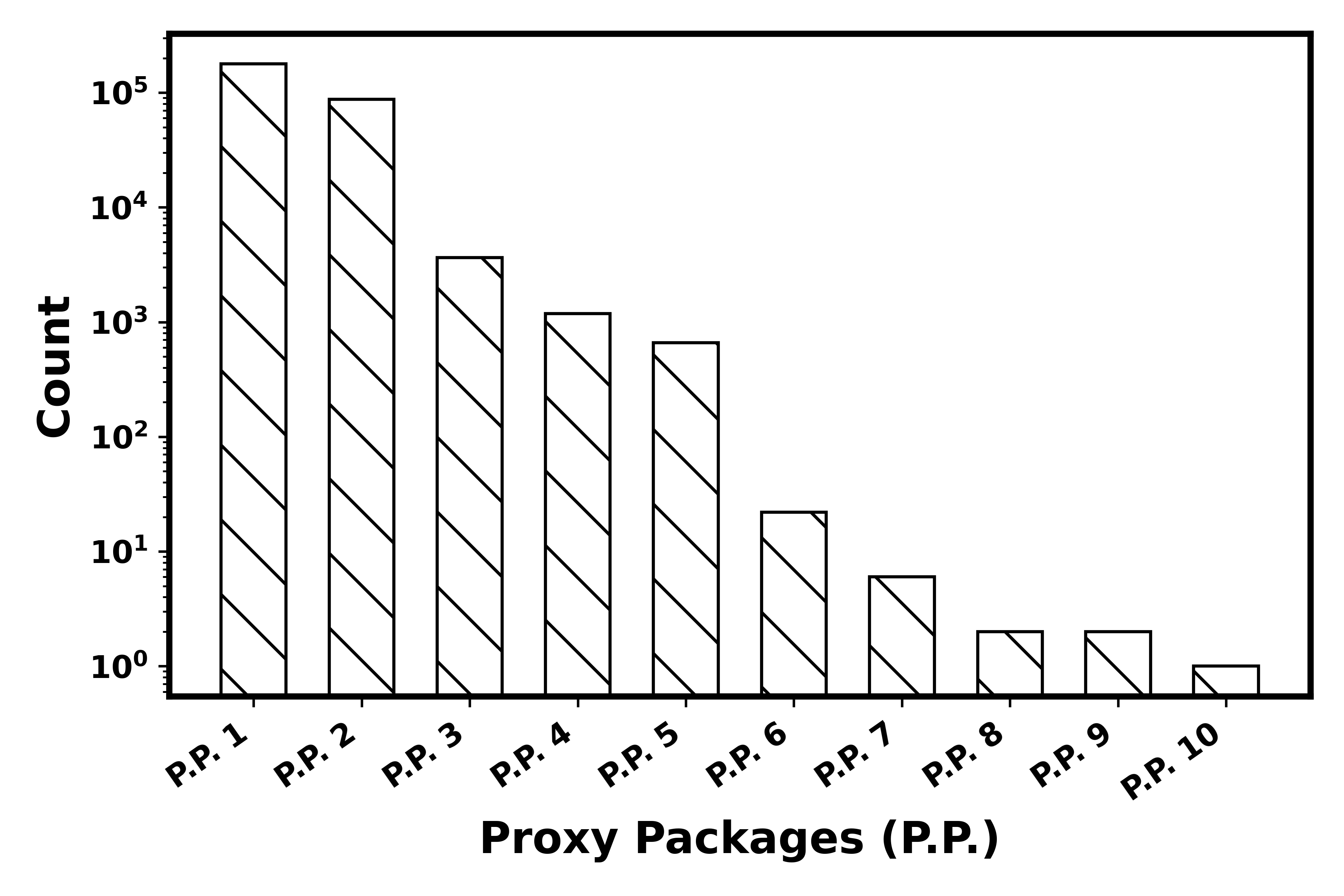}
      \label{fig:proxy_freq}
    \end{subfigure}
\caption{Frequency of VPN (only top-20) and proxy packages in validated APKs.}
\label{fig:vpn_proxy_freq}
\end{figure}

\subsection{MatchScope Results}
\label{sec:matchscope_proxy_sec}

\begin{table}[!t] 
\centering
\footnotesize{
\caption{Matching proxy APK codes to their original APKs.}
    \label{table:matchscope_proxy}
    \begin{tabular}{p{3.0cm}| R{2cm} | R{1.8cm}}
    
        \hline
        \textbf{Proxy Groups} & \textbf{Matched APKs} & \textbf{Total APKs} \\ \hline
         com.giamping.socks5, net.typeblog.socks, com.didsoft.myiphide, io.oxylabs.proxymanager	& 0 & 0 \\ \hline
         com.purevpn.proxy & 0 &	1 \\ \hline
         org.apache.http	& 684 & 78562 \\ \hline
         libv2ray, okhttp3.internal & 879 & 143519 \\ \hline
         io.netty.handler & 90 & 1010 \\ \hline
         org.littleshoot & 0 & 21 \\ \hline
         com.huawei.hms & 26 & 3150 \\ \hline

        \end{tabular}        
    }
\end{table}

\subsection{List of VPN packages \& Corresponding APK Downloads}
\label{sec:vpn_pkg}
\begin{enumerate}[noitemsep,nolistsep,leftmargin=0pt, labelindent=\parindent, listparindent=\parindent, labelwidth=0pt, itemindent=!]
    \item tun.proxy, 500+
    \item com.v2ray.ang, 40.5k
    \item com.zzy.vpnservicedemo, 36
    \item com.example.android.toyvpn, 18
    \item com.moduscreate.vpn.study, 9
    \item app.rdtunnel.pro, 100k+
    \item com.buzz.vpn, 617 
    \item com.nordvpn.android, 50M+
    \item com.fast.free.unblock.secure.vpn, RMP\footnote{RMP: Removed From Playstore}
    \item free.vpn.unblock.proxy.turbovpn, 100M+
    \item com.jrzheng.supervpnfree, RMP
    \item com.cloudflare.onedotonedotonedotone, RMP
    \item com.fast.free.unblock.thunder.vpn, RMP
    \item free.vpn.unblock.proxy.turbovpn.lite, 50M+
    \item octohide.vpn, 10M+
    \item com.expressvpn.vpn, 10M+
    \item com.free.vpn.super.hotspot.open, 100M+
    \item ch.protonvpn.android, 50M+
    \item com.windscribe.vpn, 10M+
    \item com.pawxy.browser, 1M+
    \item com.surfshark, 10M+
    \item com.security.xvpn.z35kb, RMP
    \item de.blinkt.openvpn, 10M+
    \item net.ivpn.client, 500k+
    \item com.wireguard.android, 5M+
    \item se.leap.riseupvpn, 500k+
    \item com.zerotier.one, 1M+
    \item io.nekohasekai.sfa, 100k+
    \item com.psiphon3, 50M+
    \item kittoku.osc, 100k+
    \item org.calyxinstitute.vpn, NF\footnote{NF: Not Found}
    \item free.vpn.unblock.proxy.opensource, NF
    \item com.avg.android.vpn, 5M+
    \item com.gaditek.purevpnics, 5M+
    \item tech.hexa, 5M+
    \item com.symantec.securewifi, 10M+
    \item com.softek.xnordvpnpro, 1k+
    \item com.lazycoder.cakevpn, 387
    \item com.vpn.powervpn2, 10M+
    \item com.cybernetvpn.cybernetvpn, 34
    \item com.frogobox.viprox, 21
    \item asia.buzz.freevpn, 500k+
    \item com.gaditek.purevpnics, 5M+
    \item com.proxidize.legacy.android, 306
    \item com.k3.k3pler, NF
    \item com.vesvault.vesmail, 11
    \item com.ashrafi.webi, 27
    \item com.speedy.vpn, 50M+
    \item free.vpn.unblock.proxy.vpn.master.pro, 100M+
\end{enumerate}

\subsection{List of Proxy Packages \& Corresponding APK Downloads}
\label{sec:proxy_pkg}

\begin{enumerate}[noitemsep,nolistsep,leftmargin=0pt, labelindent=\parindent, listparindent=\parindent, labelwidth=0pt, itemindent=!]
    \item libv2ray, 100k+
    \item de.blinkt.openvpn, 10M+
    \item com.giamping.socks5, 100k+
    \item io.netty.handler.proxy, 1M+
    \item org.littleshoot.proxy.impl, 1M+
    \item com.purevpn.proxy.core, RMP
    \item com.purevpn.proxy, RMP
    \item okhttp3.internal.connection, NF
    \item io.oxylabs.proxy, 100k+
    \item io.oxylabs.proxymanager, NF 
    \item org.apache.http, RMP
    \item com.didsoft.myiphide, 500k+ 
    \item net.typeblog.socks, 1M+
    \item com.huawei.hms.ads.identifier, RMP
    \item com.gorillasoftware.everyproxy.service.http, 1M+
    
\end{enumerate}

\subsection{Frequency of VPN and Proxy packages}
\label{sec:package_level_stats}
We present the count of packages in statically validated APKs with the help of \Cref{fig:vpn_proxy_freq}. We list the corresponding package names as follows -

For VPN - \hash{com.qihoo.appstore.installsecurity.vpn.LocalVPNService, 
de.blinkt.openvpn.core.OpenVPNService, org.strongswan.android.logic.CharonVpnService, 
com.sogou.androidtool.service.SgToolVpnService, net.openvpn.openvpn.OpenVPNService, 
com.anchorfree.vpnsdk.vpnservice.AFVpnService, com.tencent.pangu.utils.vpn.LocalVPNService, org.bitvise.SSHTunnelService, unified.vpn.sdk.AFVpnService, com.wireguard.android.backend.GoBackend\$VpnService, 
com.quickbird.mini.vpn.vpn.LocalVpnService, com.apld.av.vs, com.qihoo360.mobilesafe.vpn.VpnCoreService, com.v2ray.ang.service.V2RayVpnService, com.anchorfree.hydrasdk.vpnservice.AFVpnService, spt.w0pw0p.vpnlib.core.OpenVPNService
com.technore.tunnel.service.OpenVPNService, eu.faircode.netguard.ServiceSinkhole, 
com.slipkprojects.ultrasshservice.tunnel.vpn.TunnelVpnService, org.hola.vpn\_svc}

For proxy - \hash{okhttp3.internal.proxy, org.apache.http, com.huawei.hms.ads.identifier, io.netty.handler.proxy, libv2ray, org.littleshoot.proxy.impl, net.typeblog.socks, com.didsoft.myiphide, com.giamping.socks5, com.purevpn.proxy.core}

\input{sections/imp_details}

\subsection{APK Instrumentation and Execution}
\label{sec:apk_instrumentation}

To verify the usage of flagged libraries in VPN and proxy categories, we instrumented and executed a few flagged APKs and observed their runtime \texttt{logcat} results. 
We instrumented 30 flagged APKs from the proxy category and 20 flagged APKs from the VPN category using \texttt{AndroLog} tool \cite{samhi2024androlog}. Upon inspecting the logs, we observed that various library functions are being called at runtime. For example, in a VPN APK (\hash{208e8183dd41a02c5c0d2cc1b3154fe2dcaca400522ba5a178e2814cc7862124}) flagged with \texttt{unified.vpn.sdk.AFVpnService}, it populated it's class objects, and called methods related to the VPN connection initiation \textit{e.g.,} \texttt{android.net.VpnService: void onCreate()}. 

%% file: sections/ethics.tex
\subsection{Ethics Considerations}
\label{sec:ethics}
Our dynamic validation step require installation and execution of malware APKs on mobile phones so that we can monitor live CC establishments. 
Thus we carefully planned our experiments using Belmont's report \cite{beauchamp2008belmont}. We submitted a detailed IRB proposal to the institute's board. However, we were exempted from it since as per our institute's regulations, studies with no human subject involvement do not require approval. 
Still, we isolated the phones from our network so that the malware discovered none (vulnerable) to infect. The phones have been procured commercially, and are dedicated to our research. 
Further, to minimize the risk, we did not leave any malware running for too long to minimize the levels of exposure on the Internet.

%% file: sections/hyphanet.tex
\subsection{Other Anonymity Networks}
\label{sec:other_anonymity}

Apart from Tor and I2P, there are other anonymity networks such as Hyphanet (earlier Freenet), GNUNet, ZeroNet and Hopr. Among these, only Hyphanet has released an Android APK.
\subsubsection{Hyphanet/Freenet}
\label{sec:hyphanet}

Hyphanet is a peer-to-peer network proposed in 2001 \cite{clarke2001freenet}. 
We analyzed Hyphanet's APK in detail and constructed its static validation rules as follows:

\noindent \textbf{Static validation rules:} To statically check if a given APK integrates Hyphanet or not, we check for the presence of files having certain keys in them.
\begin{enumerate}
    \item Configuration files (\textit{e.g.,} \textit{freenet.ini}) with keys - \hash{console.enabled}, \hash{fproxy.enabled}, \hash{fproxy.port}, \hash{fproxy.bindTo}, \hash{node.listenPort}, \hash{node.storeSize}, \hash{node.storeType}, \hash{node.bindTo}, \hash{node.opennet.enabled}, \hash{node.opennet.listenPort}, \hash{node.opennet.bindTo}.
    \item Information about seed nodes for Freenet in files (\textit{e.g.,} \textit{seednodes.fref}) with keys -  identity=, location=, opennet=, sigP256=, physical.udp=, End.
\end{enumerate}

Application of the above rules on malware APKs from AndroZoo in \Cref{table:initial_dataset} does not flag any malware APK apart from its official one, which we are not considering as a part of our dataset.

\subsubsection{Possible reasoning for non-usage in malware}
\label{sec:no_hyphanet_reason}
There are a number of reasons why malware creators would not prefer to integrate Hyphanet in their APKs. The first one is its non-real-time communication nature, where data may take time to propagate to the network nodes. Secondly, it is geared towards working as an anonymous distributed file sharing solution, lacking the service-based models considered by other CCs like Tor, VPNs, and proxies, making it less amenable for hiding C2/bot communication.
The third reason could be the absence of support, otherwise available for projects like Tor, that are maintained and
upgraded for multiple integration methods.

%% file: sections/sota_I_table.tex
\begin{table*}[t] 
\centering
\footnotesize{
    \caption{Traffic detection results on SOTA-I \cite{barradas2024extending}. FPR/FNR (\%): Shows wrong prediction percentage.}
    \begin{tabular}{p{1.2cm}| R{0.7cm}|p{1.5cm}|p{1.3cm}|p{1.3cm}| R{0.7cm}|p{1.5cm}|p{1.3cm}|p{1.3cm} }
    
        \hline
          & \multicolumn{4}{c|}{\textbf{Benign APKs}}  & \multicolumn{4}{c}{\textbf{Malware APKs}} \\ \cline{2-9}  

        \textbf{Covert} & & \textbf{C-behav-dt} & \textbf{C-cert-rf} & \textbf{C-cert-dt} & & \textbf{C-behav-dt} & \textbf{C-cert-rf} & \textbf{C-cert-dt} \\ 
        
         \textbf{Channel} & \textbf{Flows} & \textbf{FPR (\%)} &  \textbf{FPR (\%)} & \textbf{FPR (\%)} & 
         \textbf{Flows} & \textbf{FNR (\%)} & \textbf{FNR (\%)} & \textbf{FNR (\%)} \\ \hline \hline

          Tor & 62 & 3.22 & 0.00 & 0.00  & 58 & 94.82 & 100.0 & 100.0 \\ \hline
          VPN & 4460 & 2.06 & 0.00 & 0.00 & 642 & 96.72 & 100.0 & 100.0  \\ \hline
          Proxy & 908 & 1.76 & 0.00 & 0.00  & 1687 & 97.09 & 100.0 & 100.0 \\ \hline
          TorPT & 29 & 0.00 & 0.00 & 0.00 & 568 & 97.88 & 100.0 & 100.0 \\ \hline
          I2P & 9 & 11.11 & 0.00 & 0.00 & 2623 & 95.50 & 100.0 & 100.0 \\ \hline

    \end{tabular}     
    }       

    \label{table:sota_I_results}
\end{table*}

%% file: sections/imp_details.tex
\subsection{Implementation Details}
\label{sec:impl}

We used machines running Ubuntu Linux, having atleast 16GB RAM and 6C/12T CPUs.
We used VirusTotal APIv3.0 to download VT reports and used API keys provided by \texttt{AndroZoo} to download APKs. 
In total, we downloaded 3.6M VT reports and 288k APKs which amounts to $\approx$ 100GB and $\approx$ 7TB, respectively. For malware labeling, we used \texttt{AVClass v2.8.10}. For static analysis, we used \texttt{Androguard v3.4.0a1}. 
For dynamic analysis, we used Google Pixel-5 phones. For packet analysis we used \texttt{nDPI v4.12.0}, \texttt{dpkt v1.9.8} and \texttt{Zeek v6.0.4}.